\documentclass[preprint,12pt,times]{elsarticle}

\usepackage{amssymb,amsmath}
\usepackage{hyperref,url,cleveref}
\usepackage{xcolor}
\usepackage{xspace}
\usepackage{mdframed}
\usepackage{graphicx}
\usepackage{csquotes}
\usepackage{multirow}
\usepackage[table]{xcolor}

\journal{Information and Software Technology}

\newcommand{\eg}{e.g.,\xspace}
\newcommand{\ie}{i.e.,\xspace}

\newcommand{\aka}{a.k.a.\xspace}

\definecolor{headergray}{gray}{0.7}
\definecolor{rowgray}{gray}{0.88}

\begin{document}

\begin{frontmatter}

\title{Smells and Refactorings for Software Environmental Sustainability: A Systematic Literature Review} 

\author{Ornela Danushi}
\author{Flavio Sampirisi}
\author{Jacopo Soldani} 
\author{Stefano Forti} 
\author{Antonio Brogi} 

\affiliation{organization={Department of Computer Science, University of Pisa},
            city={Pisa},
            postcode={56127}, 
            country={Italy}
            }

\begin{abstract}
\noindent \textit{Context:} 
The environmental impact of ICT is constantly growing. 
Enhancing its environmental sustainability requires dealing with energy-hungry software systems, by possibly reducing their footprint.

\noindent \textit{Objective:} 
To better frame the problem and its persistent nature, software can be viewed as carrying structural inefficiencies regardless of how advanced its functionality may be.
Identifying these inefficiencies as \enquote{green smells} provides a basis for refactoring code to mitigate its ecological impact.
We aim at identifying well-known green smells impacting environmental sustainable software and map them to corresponding refactoring aimed at reducing the impact therein.

\noindent \textit{Method:} 
In this article, we conduct a systematic literature review on the articles that discuss green smells and their possible refactorings. 
We analyse and classify 66 primary studies published up to March 2026. 

\noindent \textit{Results:} 
We identify 20 smells impacting environmental sustainability of software. Each smell is associated with one or more refactoring techniques known to mitigate their negative effects on the environmental sustainability of software.

\noindent \textit{Conclusions:} 
The literature witnesses the interest in resolving the green smells affecting software systems.
The catalogued green smells and the corresponding refactorings can provide pragmatic value for practitioners in their daily development workflows, while also offering researchers a starting point to establish new research directions in sustainable software engineering.
\end{abstract}


\end{frontmatter}




\section{Introduction}
\label{sec:introduction}

\noindent
In recent years, Information and Communication Technology (ICT) has grown significantly in both scale and complexity becoming ubiquitous in modern society (\eg technological platforms, software systems, and hardware devices), thus also increasing its footprint on the environment \cite{ID_5}.
Given that the ICT sector currently accounts for approximately 2\% of global carbon emissions, and that this is estimated to rise \cite{Jukka}, it is becoming important to tackle its environmental sustainability development.

Nowadays, enhancing the environmental sustainability of ICT not only requires dealing with hardware efficiency improvements, but also requires enhancing the energy consumption of the software itself, possibly identifying and resolving sources of inefficiencies/energy drain \cite{54_Kern2013_GreenSA}.
Although a substantial body of literature addresses environmentally sustainable software engineering, different pieces of literature deal with different aspects of sustainability. 
Software \textit{per se} carries structural inefficiencies that negatively impact energy consumption throughout its entire lifecycle; from design and development to execution and retirement \cite{Danushi2025_SLRCarbonEfficientSW,GSF2021_10Recs}.
We define such inefficiencies as \textit{green smells}, whose identification in practice remains challenging \cite{DBLP:journals/software/VerdecchiaLEVE21}. 
This underscores the need to support developers on analysing the existing software through domain-agnostic green smell detection.
Moreover, the path toward sustainability is further complicated by uncertainty regarding the precise energy footprint associated with specific green smells, as well as the trade-off between the upfront development effort required and the long-term environmental return on investment \cite{GSF2021_computation}.
This implies the need to support developers in applying \textit{refactorings} that can resolve, or at least mitigate, the negative impact of green smells, thus enhancing the environmental sustainability of affected software systems. 

At the same time, the currently available knowledge on developing environmentally sustainable software is scattered across a considerable amount of literature, posing challenges for both researchers seeking to delineate new research directions and practitioners daily needing to develop sustainable software. To address these gaps, this article tries to organise the scattered knowledge on environmentally sustainable software, by answering to the following research questions:

\begin{description}
    \item[RQ$_1$] Are there well-recognised \enquote{green smells} indicating possible issues for the environmental sustainability of existing software?
    \item[RQ$_2$] How to refactor software to mitigate the effects of green smells therein?
\end{description}

A smell can be observed in an existing software application, being it the effect of a bad decision (though often unintentional) while designing or developing such application. Such bad decision may negatively impact on the overall system quality \cite{Garcia}. The effects of smells can be mitigated by refactoring the internal structure of an application or the functionalities of its internal components, while at the same time not changing the functionalities it offers to its clients. Even if applying a refactoring requires some development efforts, it is known that refactorings mitigate the effects of smells will most probably improve the overall system quality \cite{Garcia}. Here, we target the environmental sustainability of software, from which the focus on so-called \enquote{green smells} and their possible refactorings.
Additionally, we use the term refactoring in a broader sense with respect to how defined by Fowler \cite{Fowler_refactoring} to include not only traditional source-code restructuring, but also design and configuration choices (\eg changing programming language paradigms or optimizing event rates) that directly address and mitigate the energy overhead induced by the identified green smell.



To answer to our research questions, in this article we report on what is being said by researchers about known green smells and refactorings enabling to mitigate their effects.
We indeed systematically analysed the available peer-reviewed literature on environmentally sustainable software to elicit the well-known green smells, which may increase the environmental impact of software applications.
We also elicited the refactorings proposed by researchers, which are known to mitigate the effects of such green smells.
In particular, we conducted a systematic literature review (SLR) of peer-reviewed articles on environmentally sustainable software engineering, which allowed us to map the smells that negatively affect software sustainability to the refactorings known to mitigate such effects. Following established guidelines for SLRs in software engineering \cite{Petersen2008_SMSGuidelines,Petersen2015_SMSGuidelinesUpdate}, we first identify a corpus of 66 primary studies published up to March 2026.
We then analyse and classify these studies, by eliciting 20 green smells discussed therein, as well as 33 refactorings known to mitigate the negative impact of such green smells on the environmental sustainability of software. 

Overall, our SLR seeks to provide an accurate mapping of existing green smells and their corresponding refactorings. By raising awareness on green smells and refactorings, we aim at helping developers to refactor their existing software to possibly reduce its environmental impact. Our ultimate goal is indeed supporting researchers and developers in enhancing the environmental sustainability of ICT from a software engineering perspective.

\medskip\noindent The rest of this article is organised as follows.  
\Cref{sec:design} describes the design of the SLR process.
\Cref{sec:classification} introduces the elicited green smells and their corresponding refactorings.
\Cref{sec:ttv,sec:related} discuss the possible threats to the validity of our SLR and related work, respectively.
Finally, \Cref{sec:conclusions} draws some concluding remarks and discusses future research directions.

\section{Research Design}
\label{sec:design}

\begin{figure}
    \centering
ss    \includegraphics[width=.95\textwidth]{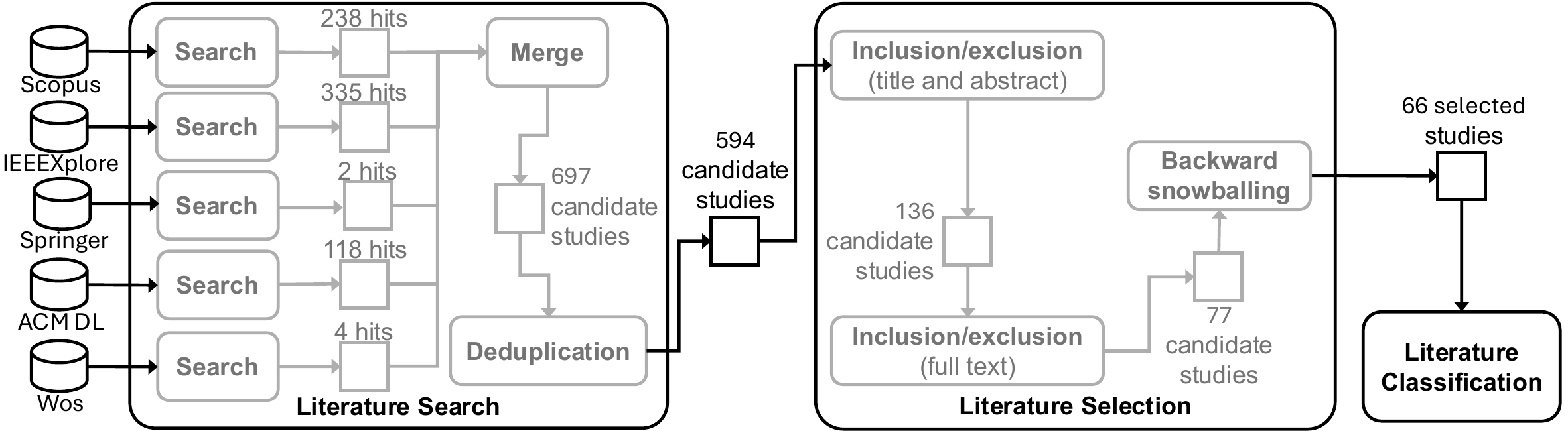} 
    \caption{SLR process for identifying and classifying existing literature of interest.} 
    \label{fig:design:workflow}
\end{figure}

\noindent
The objective of this SLR is to identify the green smells denoting possible sources of environmental impact in software, as well as the refactorings enabling to mitigate the effects of such smells.
With the objective of capturing the state of the art in the field, we searched and classified the white literature on the topic by following the process sketched in \Cref{fig:design:workflow}. 

Following established guidelines for SLRs in software engineering \cite{Petersen2008_SMSGuidelines,Petersen2015_SMSGuidelinesUpdate}, we structured the search strings guided by our main research objective, namely to overview the green smells negatively impacting on the environmental sustainability of software, together with their corresponding refactorings.
Specifically, we designed the search strings using the PICO (\textit{Population, Intervention, Comparison, Outcome}) terms \cite{Kitchenham2007_GuidelinesSLRinSE} deriving search keywords taken from each aspect of our research objective. 
We anyhow decided -- differently from what recommended by \cite{Petersen2008_SMSGuidelines,Petersen2015_SMSGuidelinesUpdate} -- to not restrict our focus to specific outcomes or experimental designs.
Instead, we aimed at providing a broader overview of the known green smells and refactorings. 
Restricting our study to given types of research settings could have introduced bias, as certain smells or refactorings might be over-/under-represented in particular study designs.
Consequently, our search string was formalised as:

\begin{footnotesize}
    \begin{align}
        & (\texttt{software*}) \wedge \\
        & (\texttt{green} \vee \texttt{sustainab*} \vee \texttt{carbon} \vee \texttt{CO2}) \wedge \\
        & (\texttt{smell*} \vee \texttt{antipattern*} \vee \texttt{bad practice* } \vee &\notag\\
        &   \texttt{pitfall*} \vee \texttt{refactor*} \vee \texttt{reengineer*})
    \end{align}
\end{footnotesize}
\noindent
(where “{\tt\small *}” matches lexically related terms, \eg plurals, conjugations, and 
synonyms).
The search string is 
the conjunction of three groups, aimed at identifying the existing efforts (1)~considering software, (2)~dealing with its environmental sustainability, and (3)~focussing on refactoring bad smells in this respect.

We used the above described string to retrieve white literature published up to March 2026, when this research began.
The search was carried out in the following indexing databases: ACM Digital Library, IEEE Xplore, Scopus, Web of Science, and SpringerLink.
We retrieved a list of 697 candidate studies, which -- after checking for duplicates -- were reduced to 594 unique candidate studies.

We then refined the list of candidate studies by relying on the inclusion and exclusion criteria listed in \Cref{tab:design:ie-criteria}.
In particular, criteria $i_1$ and $i_2$ enabled focussing on studies targeting environmentally sustainable software design and/or development. 
Criteria $i_3$ and $i_4$ further narrow the scope to studies also discussing at least one green smell and one possible refactoring for mitigatng green smell effects.
For the exclusion criteria, $e_1$, $e_2$, and $e_3$ restricted the selection to English-language, peer-reviewed literature accessible through our institution.
Finally, $e_4$, $e_5$, and $e_6$ excluded non-primary studies (\eg paper collections, secondary studies) as well as duplicate or shorter versions of already included papers.
A candidate study was selected if satisfying \textit{all} the inclusion criteria, whereas it was excluded if satisfying \textit{any} of the exclusion criteria.
\begin{table}[t]
    \caption{Inclusion and exclusion criteria.}
    \label{tab:design:ie-criteria}
    
\scriptsize
\begin{tabular}{ccp{.78\textwidth}}
    \hline 
    \rowcolor{headergray}
    \textbf{Criteria} & \textbf{Type} & \textbf{Description} \\
    \hline
    $i_1$ & Inclusion & The study explicitly targets the environmental sustainability of software \\ 
    \rowcolor{rowgray}
    $i_2$ & Inclusion & The study targets the design and/or development of software \\
    $i_3$ & Inclusion & The study discusses at least one smell/antipattern/issue for the environmental sustainability  \\ 
    \rowcolor{rowgray}
    $i_4$ & Inclusion & The study discusses at least one refactoring/solution for enhancing the environmental sustainability of software, even if the referring smell is not explicitly mentioned \\            
    $e_1$ & Exclusion & The study is not written in English  \\
    \rowcolor{rowgray}
    $e_2$ & Exclusion & The study is not peer-reviewed \\
    $e_3$ & Exclusion & The study is not accessible from our institution \\ 
    \rowcolor{rowgray}
    $e_4$ & Exclusion & The study is an edited volume collecting multiple research papers (\eg conference proceedings or edited books) rather than a single, cohesive research work. \\ 
    $e_5$ & Exclusion & The study is a secondary study (\eg a survey or SLR) \\ 
    \rowcolor{rowgray}
    $e_6$ & Exclusion & The study is a duplicate or shorter version of already included publications \\ 
    \hline 
\end{tabular}

\end{table}

The inclusion/exclusion criteria were applied in two consecutive steps.
In both steps, two authors independently evaluated each candidate study for inclusion or exclusion.
Any disagreements were resolved by a third author who had not been involved in the initial selection.
In the first step, the selection was based on assessing the inclusion/exclusion criteria over the title and abstract of candidate studies, which reduced the list of candidate studies to 136 candidate studies with an inter-rater agreement of 87.81\%.\footnote{We measured the inter-rater agreement on the inclusion/exclusion coding by exploiting the Krippendorff-$\alpha$ coefficient~\cite{Krippendorff2004_ContentAnalysis}, which classifies a coding as good if the value is greater or equal to 80\%.}
This evaluation process was then repeated on the full text of the remaining 136 candidate studies, which further narrowed the list to 32 candidate studies with an inter-rater agreement of 99.26\%.

The list of candidate studies was completed by running backward snowballing \cite{Wohlin2014_GuidelinesSnowballing}, which resulted in 77 additional candidate studies.
These were again classified for inclusion/exclusion by repeating the same process as described above and by directly considering their full read. 
As a result, 34 additional studies were selected, reaching a total of 66 studies to be considered by our SLR.
The final list of selected studies is shown in \Cref{tab:design:selected-studies}.

\begin{table}
    \caption{List of selected studies, sorted by reference number, with the last two columns including the publication type (\ie \textit{C=conference paper}, \textit{J=journal article}) and year.}
    \label{tab:design:selected-studies}
    \tiny
\begin{tabular}{clcc}
    \hline
    \rowcolor{headergray}
    \textbf{Ref.} &\textbf{Author(s). \textit{Title}} &{Type} &\textbf{Year} \\
        \hline
\cite{ID_A42} & Agolli et al. \textit{Investigating Decreasing Energy Usage in Mobile Apps via Indistinguishable Color Changes} &C &2017 \\ 
    \rowcolor{rowgray}
\cite{ID_A47} & Alam et al. \textit{Energy optimization in android applications through wakelock placement} &C &2014 \\ 
\cite{ID_A33} & Anwar et al. \textit{Evaluating the impact of code smell refactoring on the energy consumption of Android apps} &C &2019 \\ 
    \rowcolor{rowgray}
\cite{ID_A8} & Ayala et al. \textit{An energy efficiency study of web-based communication in android phones} &J &2019 \\    
\cite{ID_A57} & Banerjee and Roychoudhury. \textit{Automated re-factoring of android apps to enhance energy-efficiency} &C & 2016 \\ 
    \rowcolor{rowgray}
\cite{ID_A31} & Carette et al. \textit{Investigating the energy impact of Android smells} &C & 2017 \\     
\cite{ID_60} & Chowdhury et al. \textit{GreenBundle: An Empirical Study on the Energy Impact of Bundled Processing} &C &2019 \\ 
    \rowcolor{rowgray}
\cite{ID_A55} & Cohen et al. \textit{Energy types} &C &2012 \\ 
\cite{ID_A19} & Connolly Bree et al. \textit{Inheritance versus delegation: Which is more energy efficient?} &C & 2020 \\ 
    \rowcolor{rowgray}
\cite{ID_21} & Connolly Bree et al. \textit{Automated Refactoring for Energy-Aware Software} &C &2021 \\ 
\cite{ID_99} & Connolly Bree et al. \textit{Removing Decorator to Improve Energy Efficiency} &C &2022 \\ 
    \rowcolor{rowgray}
\cite{ID_46} & Connolly Bree et al. \textit{Energy efficiency of the Visitor Pattern: contrasting Java and C++ implementations} &J &2023 \\ 
\cite{ID_3} & Cordero et al. \textit{A First Approach on Legacy System Energy Consumption Measurement} &C & 2015 \\ 
    \rowcolor{rowgray}
\cite{ID_A28} & Couto et al. \textit{Energy refactorings for Android in the large and in the wild} & C & 2020 \\     
\cite{ID_88} & Cruz and Abreu. \textit{Performance-Based Guidelines for Energy Efficient Mobile Applications} &C &2017 \\ 
    \rowcolor{rowgray}
\cite{ID_A27} & Cruz and Abreu. \textit{Using automatic refactoring to improve energy efficiency of Android apps} &J &2018 \\ 
\cite{ID_77} & Cruz et al. \textit{Leafactor: Improving Energy Efficiency of Android Apps via Automatic Refactoring} &C &2017 \\ 
    \rowcolor{rowgray}
\cite{ID_A2} & Cruz et al. \textit{Innovating for Tomorrow: The convergence of SE and Green AI} &J &202) \\ 
\cite{ID_10} & Dhaka et al. \textit{An Empirical Investigation into Code Smell Elimination Sequences for Energy Efficient Software} &C & 2016 \\ 
    \rowcolor{rowgray}
\cite{ID_94} & Fawad et al. \textit{Refactoring Android Source Code Smells From Android Applications} &J &2025 \\     
\cite{ID_A54} & Gattal et al. \textit{Exploiting the progress of OO refactoring tools with Android code smells: RAndroid, a plugin for Android studio} &C &2021 \\ 
    \rowcolor{rowgray}
\cite{ID_18} & Giavrimis et al. \textit{Artemis AI: Multi-LLM Framework for Code Optimisation} &C &2025 \\ 
\cite{ID_49} & Le Goaër. \textit{Enforcing Green Code With Android Lint} &C & 2020 \\ 
    \rowcolor{rowgray}
\cite{ID_110} & Gurung et al. \textit{Static Code Analysis for Reducing Energy Code Smells in Different Loop Types: A Case Study in Java} &C &2024 \\
\cite{ID_A4} & Hasan et al. \textit{Energy profiles of java collections classes} &J &2016 \\
    \rowcolor{rowgray}
\cite{ID_126} & Hecktor and Sachweh. \textit{Towards a Conceptual Framework for Greener Software Systems} &C & 2025 \\ 
\cite{ID_A29} & Iannone et al. \textit{Refactoring Android-specific energy smells: A plugin for Android Studio} &C &2020 \\
    \rowcolor{rowgray}
\cite{ID_129} & Imran et al. \textit{Towards Sustainable Cloud Software Systems through Energy-Aware Code Smell Refactoring} &C & 2024 \\     
\cite{ID_48} & Jolly et al. \textit{Energy-Aware Computing: Cutting CPU Cycles through Intelligent Thread Management and Code Smell Optimization} &C &2025 \\ 
    \rowcolor{rowgray}
\cite{ID_25} & Kim et al. \textit{Code refactoring techniques for reducing energy consumption in embedded computing environment} &J & 2016 \\ 
\cite{ID_A70} & Lano et al. \textit{Software modelling for sustainable software engineering} &C & 2024 \\ 
    \rowcolor{rowgray}
\cite{ID_A69} & Lano et al. \textit{Design patterns for software sustainability} &J & 2024 \\
\cite{ID_113} & Lano et al. \textit{Sustainability Implications of Large Language Model use in Model-Driven Engineering} &C &2025 \\     
    \rowcolor{rowgray}
\cite{ID_A7} & Li and Halfond. \textit{An investigation into energy-saving programming practices for android smartphone application development} &J & 2014 \\
\cite{ID_A56} & Li and Halfond. \textit{Optimizing energy of http requests in android applications} &C & 2015 \\ 
    \rowcolor{rowgray}
\cite{ID_A44} & Li et al. \textit{Automated energy optimization of http requests for mobile applications} &C & 2016 \\ 
\cite{ID_A46} & Li et al. \textit{Making Web Applications More Energy Efficient for OLED Smartphones} &C & 2014 \\ 
    \rowcolor{rowgray}
\cite{ID_A48} & Lyu et al. \textit{Remove rats from your code: Automated optimization of resource inefficient database writes for mobile applications} &C & 2018 \\
\cite{ID_130} & Marinescu. \textit{Towards the impact of design flaws on the resources used by an application} &J & 2014 \\ 
    \rowcolor{rowgray}
\cite{ID_A10} & Morales et al. \textit{Earmo: An energy-aware refactoring approach for mobile apps} &C & 2018 \\ 
\cite{ID_A68} & Moreira et al. \textit{Overviewing the liveness of refactoring for energy efficiency} &C & 2020 \\ 
    \rowcolor{rowgray}
\cite{ID_124} & Nagy et al. \textit{Tools supporting green computing in Erlang} &C & 2019 \\ 
\cite{ID_A1} & Naumann et al. \textit{The GREENSOFT Model: A reference model for green and sustainable software and its engineering} &J & 2011 \\
    \rowcolor{rowgray}
\cite{ID_5} & O'Dea. \textit{A Measurement-Driven Approach to Enhancing Sustainability in Microservice Architectures} &C & 2025 \\ 
\cite{ID_116} & Ournani et al. \textit{Tales from the code \#1: The effective impact of code refactorings on software energy consumption} &C & 2021 \\
    \rowcolor{rowgray}
\cite{ID_A30} & Palomba et al. \textit{Lightweight detection of Android-specific code smells: The aDoctor project} &C & 2017 \\     
\cite{ID_86} & Palomba et al. \textit{On the impact of code smells on the energy consumption of mobile applications} &J & 2019 \\
    \rowcolor{rowgray}
\cite{ID_A3} & Park et al. \textit{Investigation for Software Power Consumption of Code Refactoring Techniques} &C & 2014 \\ 
\cite{ID_120} & Pereira et al. \textit{The Influence of the Java Collection Framework on Overall Energy Consumption} &C &2016 \\ 
    \rowcolor{rowgray}
\cite{ID_75} & Pereira et al. \textit{jStanley: Placing a Green Thumb on Java Collections} &C & 2018 \\ 
\cite{ID_47} & Rani et al. \textit{Energy Patterns for Web: An Exploratory Study} &C &2024 \\ 
    \rowcolor{rowgray}
\cite{ID_A35} & Reeshti et al. \textit{Code Smell Refactoring for Energy Optimization of Android Apps} &C &2021 \\     
\cite{ID_38} & Ribeiro et al. \textit{EcoAndroid: An Android Studio Plugin for Developing Energy-Efficient Java Mobile Apps} &C & 2021 \\
    \rowcolor{rowgray}
\cite{ID_A53} & Saborido et al. \textit{Getting the most from map data structures in Android} &J &2018 \\ 
\cite{ID_67} & Sahin et al. \textit{How do code refactorings affect energy usage?} &C &2014 \\ 
    \rowcolor{rowgray}
\cite{ID_A9} & Sahin et al. \textit{From benchmarks to real apps: Exploring the energy impacts of performance-directed changes} &J &2016 \\
\cite{ID_44} & Sanlialp et al. \textit{Energy Efficiency Analysis of Code Refactoring Techniques for Green and Sustainable Software in Portable Devices} &C &2022 \\
        \rowcolor{rowgray}
\cite{ID_59} & Sehgal et al. \textit{Green software: Refactoring approach} &J &2022 \\ 
\cite{ID_A12} & Silva et al. \textit{Evaluation of the impact of code refactoring on embedded software efficiency} &C & 2010 \\ 
    \rowcolor{rowgray}
\cite{ID_A40} & Singh et al. \textit{Impact of developer choices on energy consumption of software on servers} &C & 2015 \\
\cite{ID_A6} & Tonini et al. \textit{Analysis and evaluation of the android best practices impact on the efficiency of mobile apps} &C & 2013 \\     
    \rowcolor{rowgray}
\cite{ID_57} & Toosi. \textit{Green Software Engineering: A Dual-Perspective Overview of Stakeholder and Societal Interpretations} &C & 2025 \\
\cite{ID_A15} & Verdecchia et al. \textit{Empirical evaluation of the energy impact of refactoring code smells} &C & 2018 \\
    \rowcolor{rowgray}
\cite{ID_24} & Werner et al. \textit{Code once, Run Green: Automated Green Code Translation in Serverless Computing} &C & 2025 \\ 
\cite{ID_A24} & Wertjanz et al. \textit{Removing energy code smells with reengineering services} &C & 2012 \\ 
    \rowcolor{rowgray}
\cite{ID_100} & Yadav et al. \textit{Revolutionizing Software Development: Enhancing Quality and Performance Through Code Refactoring} &C &2024 \\
        \hline    
    \end{tabular}

\end{table}

\section{Green Smells and Refactorings}
\label{sec:classification}

\noindent
We hereafter detail the systematic approach that we followed, by describing the triangulation and inter-rater reliability assessment trials that we ran to enforce the validity of our findings.
Specifically, we adopted thematic coding \cite{thematic_coding} by annotating and labelling the selected studies with identified green smells and refactorings on environmental sustainability of software. 
This phase was completed with several iterative rounds conducted among all authors, in an effort to minimise possible observer biases.
Particularly, we obtained our catalogue of green smells and refactoring by following the guidelines for conducting systematic literature reviews in software engineering \cite{Kitchenham2007_GuidelinesSLRinSE}.
We identified the green smells by performing a first scan of the selected primary studies. 
Then, we extrapolated the refactorings directly from the selected primary studies after additional scans.
The obtained green smells and refactorings were then manually organised to obtain a catalogue.
Such catalogue underwent various iterations among the authors, followed by a final joint discussion session that resulted in the green smells and refactorings displayed in \Cref{tab:classification:taxonomy}.


\begin{table}[t]
    \caption{Catalogue of green smells and refactorings. The smells are sorted by frequency, namely by the most frequently discussed to the least frequently discussed in the selected studies. 
    }
    \label{tab:classification:taxonomy}
    
\scriptsize
\centering
\begin{tabular}{cl} 

    \hline 
    \rowcolor{headergray}
    \textbf{Green smell} & \textbf{Refactorings} \\
    \hline

Inefficient data structures 
    & Use optimised data structures \\ 
\rowcolor{rowgray}
Multiple responsibility classes 
    & Follow single responsibility principle \\

Excessive looping
    & Remove loop initialisation \\
    & Separate exit condition \\
    & Unroll nested loops \\
    & Use optimised loop \\

\rowcolor{rowgray}
Durable resource usage
    & Lock resources only when necessary \\

Heavy graphics
    & View usage optimisation \\
    & Avoid useless animation \\
    & Dark colour usage \\

\rowcolor{rowgray}
Inefficient transmission
    & Reduce payload \\
\rowcolor{rowgray}
    & Use dynamic retry delay \\

Excessive encapsulation
    & Reduce encapsulation \\

\rowcolor{rowgray}
Redundant code
    & Remove obsolete code \\
    
Excessive event rate
    & Push over pull \\
    & Reduce event rate \\
    & Batch events \\

\rowcolor{rowgray}
Excessive indirection
    & Reduce indirection \\
    
Excessive delegation
    & Reduce delegation \\

\rowcolor{rowgray}
Long function
    & Decompose function \\

Repeated computation
    & Extract local variable \\
    & Use caching \\
    & Use pre-fetching \\

\rowcolor{rowgray}
Inefficient parallelism
    & Avoid leaking threads \\
\rowcolor{rowgray}
    & Limit threads \\

Wrong variable types
    & Move global to local variables \\
    & Move static to instance variables \\

\rowcolor{rowgray}
Inefficient algorithm
    & Use optimised algorithm \\
    
Inefficient third-party APIs
    & Use energy-efficient APIs \\

\rowcolor{rowgray}
Single functioning
    & Enable emergency mode \\
    
Interpreted programming language
    & Switch to compiled language \\

\rowcolor{rowgray}
Useless recursion  
    & Use iterative solution \\

    \hline 
\end{tabular}

\end{table}

\Cref{tab:classification:selected-studies} shows the classification of all selected studies, thus providing an overview of the coverage of the green in the selected literature.
Below, we discuss each identified smell, highlighting its impact along with the corresponding refactoring techniques used for mitigation.

\begin{table}
    \caption{Classification of the selected studies based on the catalogue in \Cref{tab:classification:taxonomy}.}
    \label{tab:classification:selected-studies}
    \resizebox{\textwidth}{!}{

\begin{tabular}{|c|c|c|c|c|c|c|c|c|c|c|c|c|c|c|c|c|c|c|c|c|}
\hline

&ineff. &mult. &&durab. &&&&&excess. &&&&&&wrong &&ineff. &&inter. &\\

&data &resp. &excess. &resour. &heavy &ineff. 
&excess. &redund. &event. &excess. & excess. &long &repeat. &ineff. &var. &ineff. &third. &single 
&program. &useless\\

\textbf{Ref.} 
&struct. &classes &loop. &usage &graphics 
&trans. &encap. &code &rate &indir. &deleg. &func &comp. &parall. &types &algo. &APIs. &func. &lang. &recursion\\

\hline

\cite{ID_A42} &  &  &  &  & \checkmark &  &  &  &  &  &  &  &  &  &  &  &  &  &  &  \\ \hline
\cite{ID_A47} &  &  &  & \checkmark &  &  &  &  &  &  &  &  &  &  &  &  &  &  &  &  \\ \hline
\cite{ID_A33} &  & \checkmark &  &  &  &  &  & \checkmark &  &  &  & \checkmark &  &  & \checkmark &  &  &  &  &  \\ \hline
\cite{ID_A8} &  &  &  &  &  &  &  &  & \checkmark &  &  &  &  &  &  &  &  & \checkmark &  &  \\ \hline
\cite{ID_A57} &  &  &  & \checkmark &  &  &  &  & \checkmark &  &  &  &  &  &  &  &  &  &  &  \\ \hline
\cite{ID_A31} & \checkmark & \checkmark &  &  &  &  & \checkmark &  &  &  &  &  &  &  &  &  &  &  &  &  \\ \hline
\cite{ID_60} &  &  &  &  &  & \checkmark &  &  & \checkmark &  & \checkmark &  & \checkmark &  &  &  &  &  &  &  \\ \hline
\cite{ID_A55} &  &  &  &  &  &  &  &  &  &  &  &  &  &  &  &  &  & \checkmark &  &  \\ \hline
\cite{ID_A19} &  &  &  &  &  &  &  &  &  &  & \checkmark &  &  &  &  &  &  &  &  &  \\ \hline
\cite{ID_21} & \checkmark & \checkmark & \checkmark & \checkmark & \checkmark & \checkmark & \checkmark & \checkmark & \checkmark & \checkmark & \checkmark & \checkmark &  &  &  & \checkmark & \checkmark &  &  &  \\ \hline
\cite{ID_99} &  &  &  & \checkmark & \checkmark &  & \checkmark &  &  & \checkmark & \checkmark &  &  &  &  &  &  &  &  &  \\ \hline
\cite{ID_46} &  & \checkmark &  & \checkmark & \checkmark &  & \checkmark &  &  & \checkmark & \checkmark & \checkmark &  &  &  &  &  &  & \checkmark &  \\ \hline
\cite{ID_3} &  &  & \checkmark &  &  & \checkmark &  &  &  &  &  &  &  &  &  &  &  &  &  &  \\ \hline
\cite{ID_A28} & \checkmark & \checkmark &  & \checkmark & \checkmark &  &  &  &  & \checkmark &  &  &  & \checkmark &  &  &  &  &  &  \\ \hline
\cite{ID_88} & \checkmark &  &  &  & \checkmark & \checkmark &  &  & \checkmark &  &  &  &  & \checkmark &  &  & \checkmark & \checkmark &  &  \\ \hline
\cite{ID_A27} &  &  &  & \checkmark & \checkmark &  &  &  &  &  &  &  &  & \checkmark &  &  &  &  &  &  \\ \hline
\cite{ID_77} &  &  &  & \checkmark & \checkmark &  &  &  &  &  &  &  &  & \checkmark &  &  &  &  &  &  \\ \hline
\cite{ID_A2} &  &  &  &  &  &  &  & \checkmark & \checkmark &  &  &  &  &  &  &  &  &  &  &  \\ \hline
\cite{ID_10} &  & \checkmark &  &  &  &  &  &  &  &  &  &  &  &  &  &  &  &  &  &  \\ \hline
\cite{ID_94} & \checkmark & \checkmark & \checkmark & \checkmark &  & \checkmark & \checkmark &  & \checkmark &  &  &  &  & \checkmark &  &  &  & \checkmark &  &  \\ \hline
\cite{ID_A54} & \checkmark & \checkmark &  &  & \checkmark &  &  & \checkmark & \checkmark &  &  & \checkmark &  &  &  &  &  &  &  &  \\ \hline
\cite{ID_18} &  &  & \checkmark &  &  &  &  &  &  &  &  &  & \checkmark &  &  &  &  &  &  &  \\ \hline
\cite{ID_49} &  &  &  & \checkmark & \checkmark & \checkmark &  &  & \checkmark &  &  &  &  &  &  &  & \checkmark & \checkmark &  &  \\ \hline
\cite{ID_110} & \checkmark &  & \checkmark &  &  &  &  &  &  &  &  &  &  &  & \checkmark &  &  &  &  &  \\ \hline
\cite{ID_A4} & \checkmark &  &  &  &  &  &  &  &  &  &  &  &  &  &  & \checkmark &  &  &  &  \\ \hline
\cite{ID_126} &  &  &  &  &  & \checkmark &  &  &  &  & \checkmark &  &  &  & \checkmark &  & \checkmark &  & \checkmark &  \\ \hline
\cite{ID_A29} & \checkmark & \checkmark &  &  &  &  & \checkmark &  &  &  &  &  &  & \checkmark &  &  &  &  &  &  \\ \hline
\cite{ID_129} & \checkmark & \checkmark & \checkmark &  &  &  &  & \checkmark &  & \checkmark & \checkmark & \checkmark &  &  &  &  &  &  &  &  \\ \hline
\cite{ID_48} & \checkmark &  & \checkmark &  &  &  &  & \checkmark &  &  &  & \checkmark & \checkmark & \checkmark &  &  &  &  &  &  \\ \hline
\cite{ID_25} & \checkmark &  & \checkmark & \checkmark & \checkmark &  &  & \checkmark &  & \checkmark & \checkmark &  &  &  & \checkmark &  & \checkmark &  &  & \checkmark \\ \hline
\cite{ID_A70} & \checkmark & \checkmark & \checkmark &  &  & \checkmark &  & \checkmark &  & \checkmark &  &  & \checkmark &  &  &  &  &  &  &  \\ \hline
\cite{ID_A69} &  & \checkmark &  &  &  &  & \checkmark &  &  &  & \checkmark &  & \checkmark &  &  &  &  &  &  &  \\ \hline
\cite{ID_113} & \checkmark &  & \checkmark &  &  &  &  &  &  &  &  &  & \checkmark &  &  & \checkmark &  &  &  & \checkmark \\ \hline
\cite{ID_A7} &  &  & \checkmark &  & \checkmark & \checkmark & \checkmark &  & \checkmark & \checkmark &  &  & \checkmark &  &  & \checkmark &  &  &  &  \\ \hline
\cite{ID_A56} &  &  &  &  &  & \checkmark &  &  & \checkmark &  &  &  &  &  &  &  &  &  &  &  \\ \hline
\cite{ID_A44} &  &  &  &  &  & \checkmark &  &  &  &  &  &  &  &  &  &  &  &  &  &  \\ \hline
\cite{ID_A46} &  &  &  &  & \checkmark &  &  &  &  &  &  &  &  &  &  &  &  &  &  &  \\ \hline
\cite{ID_A48} &  &  &  &  &  &  &  &  & \checkmark &  &  &  &  &  &  &  &  &  &  &  \\ \hline
\cite{ID_130} &  & \checkmark &  &  &  &  &  &  &  & \checkmark &  &  &  &  &  &  &  &  &  &  \\ \hline
\cite{ID_A10} & \checkmark & \checkmark &  & \checkmark &  &  & \checkmark &  &  &  &  & \checkmark &  &  &  &  &  &  &  &  \\ \hline
\cite{ID_A68} & \checkmark & \checkmark &  & \checkmark & \checkmark &  & \checkmark &  &  &  &  &  &  &  &  &  &  &  &  &  \\ \hline
\cite{ID_124} & \checkmark &  &  &  &  & \checkmark &  &  &  &  & \checkmark &  &  & \checkmark &  & \checkmark &  &  &  &  \\ \hline
\cite{ID_A1} &  &  &  &  &  &  &  &  &  &  &  &  & \checkmark &  & \checkmark &  &  &  &  &  \\ \hline
\cite{ID_5} &  &  &  & \checkmark &  & \checkmark &  & \checkmark &  &  &  &  &  &  &  & \checkmark &  &  &  &  \\ \hline
\cite{ID_116} & \checkmark & \checkmark &  &  & \checkmark &  &  & \checkmark &  &  &  &  &  &  & \checkmark &  &  &  &  &  \\ \hline
\cite{ID_A30} & \checkmark & \checkmark & \checkmark &  &  & \checkmark & \checkmark &  & \checkmark &  &  &  &  & \checkmark &  &  &  & \checkmark &  &  \\ \hline
\cite{ID_86} & \checkmark & \checkmark & \checkmark &  &  & \checkmark & \checkmark &  &  &  & \checkmark & \checkmark &  & \checkmark &  &  &  &  &  &  \\ \hline
\cite{ID_A3} &  & \checkmark & \checkmark &  &  &  & \checkmark &  &  & \checkmark & \checkmark &  &  &  &  & \checkmark &  &  &  &  \\ \hline
\cite{ID_120} & \checkmark &  &  &  &  &  &  &  &  &  & \checkmark &  &  &  &  &  &  &  &  &  \\ \hline
\cite{ID_75} & \checkmark &  &  &  &  &  &  &  &  &  &  &  &  &  &  &  &  &  & \checkmark &  \\ \hline
\cite{ID_47} &  &  &  & \checkmark & \checkmark & \checkmark &  &  & \checkmark &  &  &  & \checkmark &  &  &  & \checkmark & \checkmark &  &  \\ \hline
\cite{ID_A35} &  & \checkmark &  &  &  &  &  & \checkmark &  &  &  & \checkmark &  &  & \checkmark &  &  &  &  &  \\ \hline
\cite{ID_38} &  &  &  & \checkmark & \checkmark & \checkmark &  &  & \checkmark &  &  &  & \checkmark &  &  &  &  &  &  &  \\ \hline
\cite{ID_A53} & \checkmark &  &  &  &  &  &  &  &  &  &  &  &  &  &  &  &  &  &  &  \\ \hline
\cite{ID_67} &  &  &  &  &  &  &  &  &  & \checkmark &  &  & \checkmark &  & \checkmark & \checkmark &  &  &  &  \\ \hline
\cite{ID_A9} &  & \checkmark & \checkmark & \checkmark &  &  &  &  &  &  &  &  &  &  & \checkmark &  &  &  &  &  \\ \hline
\cite{ID_44} &  &  & \checkmark &  &  &  & \checkmark &  &  & \checkmark & \checkmark & \checkmark &  &  &  &  &  &  &  &  \\ \hline
\cite{ID_59} &  & \checkmark &  &  &  &  &  & \checkmark &  &  &  & \checkmark &  &  &  &  &  &  &  &  \\ \hline
\cite{ID_A12} &  &  &  &  &  &  &  &  &  & \checkmark &  &  &  &  &  &  &  &  &  &  \\ \hline
\cite{ID_A40} &  &  &  &  &  &  &  &  &  &  &  &  &  &  &  &  & \checkmark &  &  &  \\ \hline
\cite{ID_A6} &  &  & \checkmark &  &  &  & \checkmark &  &  &  &  &  &  &  &  &  &  &  &  &  \\ \hline
\cite{ID_57} & \checkmark &  & \checkmark &  &  &  &  & \checkmark &  &  &  &  &  & \checkmark &  & \checkmark &  &  & \checkmark &  \\ \hline
\cite{ID_A15} &  & \checkmark &  &  &  &  & \checkmark & \checkmark &  &  &  & \checkmark &  &  & \checkmark &  & \checkmark &  &  &  \\ \hline
\cite{ID_24} &  &  &  &  &  &  &  &  &  &  &  &  &  &  &  &  &  &  & \checkmark &  \\ \hline
\cite{ID_A24} & \checkmark &  &  & \checkmark &  &  &  & \checkmark & \checkmark & \checkmark &  &  &  &  &  &  &  &  &  &  \\ \hline
\cite{ID_100} & \checkmark & \checkmark & \checkmark &  &  &  &  & \checkmark &  & \checkmark &  &  &  &  & \checkmark & \checkmark &  &  &  & \checkmark \\ \hline

\textbf{Tot.} &26 &24 &19 &18 &17 &17 &16 &16 &16 &15 &14 &12 &11 &11 &11 &10 &8 &7 &5 &3 \\ \hline

\end{tabular}
}
\end{table}


\subsection{Inefficient Data Structures}
\label{sec:classification:IDS}
\noindent
The \textsf{Inefficient Data Structures} smell arises when a data structure is ill-suited for its intended usage pattern \cite{ID_48, ID_A10, ID_A24}. 
Consequently, the \textsf{use optimised data structures} refactoring involves selecting more suitable data structure implementations based on the execution context \cite{ID_21, ID_124, ID_120, ID_57}. 
For instance, as the workload expands, an initial selected data structure may fail to scale efficiently with increasing data size~\cite{ID_75}.
In other scenarios, an inappropriate choice occurs when entire data structures are passed by value rather than by reference, leading to increased energy consumption; in such cases, replacing a full data structure with a pointer allows memory to be directly modified by called functions \cite{ID_25}.
Similarly, using string literals when specific operations must be performed on individual items is inefficient, and replacing them with enumerations represents a valid refactoring \cite{ID_100}.
Finally, when compute membership checks are required on data structures where such operations are costly, the smell can be resolved by replacing sequence-based types with set-based structures \cite{ID_A70, ID_113}.

Alternatively, this code smell may arise from selecting a generic data type when a more specific one would offer greater efficiency.
The most frequent instance involves mapping primitive types (\eg integers) to object wrappers using generic collections such as \textsf{HashMap} \cite{ID_94}).
To mitigate this, common refactoring involves replacing \textsf{HashMap} either with \textsf{SparseArray} \cite{ID_A54, ID_A29, ID_A30, ID_86} or \textsf{ArrayMap} \cite{ID_A31, ID_A28, ID_A10, ID_A68, ID_A53}. 
Similarly, substituting a \textsf{LinkedList} with \textsf{ArrayList} \cite{ID_A4} replaces a generic collection with a more specialised structure tailored for specific access patterns.

Additionally, resource-specific alternatives require careful management; for instance, when using \textsf{TypedArray}, developers must explicitly handle the lifecycle of singleton resources to release them properly \cite{ID_88}, or leverage pre-configured data structure to avoid synchronisation overhead \cite{ID_129}.
Finally, repeated string concatenation using the \texttt{+} operator instantiates unnecessary transient objects compared to mutable alternatives such as the Java's \textsf{StringBuilder} \cite{ID_94, ID_110, ID_116}, for which further optimised implementations are available.

\subsection{Multiple Responsibility Classes}
\label{sec:classification:MRC}
\noindent
The \textsf{Multiple Responsibility Classes} smell occurs when software is designed with a over-generic behaviour that consumes excessive energy by failing to specialise execution based on actual user needs \cite{ID_A70, ID_A69, ID_A35}. By grouping many activities into a few monolithic parts of the software, this smell exacerbates resource consumption and typically manifests through long, often complex implementations \cite{ID_21, ID_46, ID_130}. This smell spans different granularity levels, ranging from methods that ignore class properties \cite{ID_94} to variables misplaced across member classes \cite{ID_129}, and data organisation per se \cite{ID_A3}.

Its detection is particularly straightforward in object-oriented paradigms, where the misplacement of class responsibilities and poor structural organisation can be readily analysed \cite{ID_A54, ID_A10, ID_A68, ID_86}. 
Specifically, by explicitly defining and scoping class responsibilities, the refactored implementation applies the \textsf{follow single responsibility principle} mitigating the negative impacts of this smell.
For instance, by separating abstract interfaces from concrete implementations facilitates the execution of this refactoring \cite{ID_59}.

Additionally, because different parts of the software can be affected by this smell, the order and extent of refactoring must be carefully planned to maximise energy reduction \cite{ID_10}. In general, the refactoring involves splitting, extracting, and moving misplaced components to their proper locations \cite{ID_A33, ID_116, ID_A9, ID_A15, ID_100}. 
This organisation can also involve the transformation of eligible methods into static methods, as bytecode symbols are resolved at compile time rather than runtime \cite{ID_A31, ID_A28, ID_A29, ID_A30}.

\subsection{Excessive Looping}
\label{sec:classification:EL}
\noindent
The \textsf{Excessive Looping} smell denotes a flawed loop design that degrades energy efficiency. For instance, initialising a collection -- most commonly an array -- within a loop body introduces energy consumption due to repeated index increments, loop comparisons, and branching operations. 
Consequently, to mitigate these negative impacts, the \textsf{remove loop initialisation} refactoring either moves the collection initialisation outside the loop structure or caches the collection size across iterations to prevent redundant boundary condition evaluations across iterations \cite{ID_A7, ID_A9, ID_A6}.

Flawed exit conditions also introduce this smell \cite{ID_110}.
In such cases, applying the \textsf{separate exit conditions} refactoring introduces explicit break statements to handle termination cleanly \cite{ID_25, ID_100}. 
Alternatively, this smell arises from overly complex loop logic.
For instance, deeply nested loop structures \cite{ID_3, ID_120} severly impair energy efficiency, particularly during intensive operations such as matrix computations \cite{ID_94}. 
To mitigate this, the \textsf{unroll nested loops} refactoring reduces nesting depth of the implementation \cite{ID_18, ID_129, ID_48, ID_A70, ID_A3, ID_44, ID_57}.
Finally, performance bottlenecks in slow loop implementations \cite{ID_86} can be mitigated with the \textsf{use optimised loop} refactoring.
This is typically achieved either by replacing a standard \textit{for}-loop with an enhanced \textit{for-each} construct \cite{ ID_21, ID_A30} or by simplifying the operations within the loop body \cite{ID_A70,ID_113}.

\subsection{Durable Resource Usage}
\label{sec:classification:DRU}
\noindent
The \textsf{Durable Resource Usage} smell pertains to resource contention -- occurring when threads attempt to access unavailable resources \cite{ ID_5, ID_38} -- or unnecessary dependencies caused by improper resource allocations. 
To mitigate the negative impacts on energy consumption, the \textsf{lock resources only when necessary} refactoring aims at optimising the scope of resource locking \cite{ID_A47, ID_46, ID_A68}.
Specifically, when resources are improperly allocated well in advance of their actual usage -- resulting in premature binding \cite{ID_21, ID_99, ID_94, ID_A24} -- this refactoring defers allocations until execution, thereby preventing premature resource binding \cite{ID_A10}.

Alternatively, this smell can occur after resource usage due to a failure to release resources, leading to resource leaks \cite{ID_25, ID_A9, ID_A24}.
Typical examples include singleton resources that require manual closure \cite{ID_77} or failing to invoke the \textsf{close()} method on closable objects \cite{ID_94}.
The corresponding refactoring aims at releasing unnecessary resources as early as possible \cite{ID_49}.
For instance, this can be achieved by invoking a \textsf{recycle()} method to automatically release all allocated resources before a method returns \cite{ID_A28, ID_A27} or prior to application exit \cite{ID_A57}.
Conversely, when resource demand can be predicted in the near future, retaining hold of allocated resources -- counterintuitive though it may seem -- can reduce energy consumption in specific scenarios \cite{ID_47}.

\subsection{Heavy Graphics}
\label{sec:classification:HG}
\noindent
The \textsf{Heavy Graphics} smell pertains the energy-intensive processing of graphical user interface components. 
When this smell arises from display hardware characteristics, such as energy consumption induced by specific colour schemes in mobile and web interfaces \cite{ID_25}, adopting the \textsf{dark colour usage} refactoring serves an effective energy-efficient countermeasure \cite{ID_A42, ID_21, ID_99, ID_46, ID_49, ID_A7, ID_A46}.
When this smell arises from excessive animations, the \textsf{avoid useless animation} refactoring tends to use resources, graphic tasks, and pixels with moderation to avoid computations for those that are unnecessary \cite{ID_88, ID_A54, ID_47, ID_38}. 

Alternatively, inefficient view manipulation can negatively impact energy consumption due to many reasons for which the \textsf{view usage optimisation} refactoring pertains to.
Specifically, when allocations happen within the drawing logic (\aka DrawAllocation) \cite{ID_77}, move the object allocation outside the method affected by this smell and reuse such objects for each drawing operation \cite{ID_A28, ID_A27, ID_A68}.
Otherwise, when there are unused layout parameters, UI items, or layouts (\aka ObsoleteLayoutParam, UnusedResources, UselessParent, respectively) \cite{ID_77, ID_116}, the applied refactoring aims at removing them \cite{ID_A28, ID_88, ID_A27}.
Finally, when the same regions are painted more than once (\aka Overdraw) \cite{ID_A54, ID_116}, also intended for a smoother scroll in list views (\aka ViewHolder) \cite{ID_116}, the refactoring aims at removing the background of parent views \cite{ID_88} or recycle previous view processed data as much as possible \cite{ID_77, ID_A68}.

\subsection{Inefficient Transmission}
\label{sec:classification:IT}
\noindent
The \textsf{Inefficient Transmission} smell pertains to data transfers that degrade energy efficiency
during inter-service messaging within containerised environments. 
In such cases, the \textsf{reduce payload} refactoring aims at minimising communication overhead through strategies as container consolidation, optimised orchestration, strategic service placement, and decentralised request routing \cite{ID_5}. 
Alternatively, this smell can occur when accessing remote databases \cite{ID_3, ID_88, ID_94, ID_A30}. Under these circumstances, the \textsf{reduce payload} refactoring serves as a broad query optimisation strategy \cite{ID_60, ID_86}.

This smell can also occur during network infrastructure communications \cite{ID_88, ID_94, ID_49, ID_86}.
In this case, the \textsf{reduce payload} refactoring targets two different mechanisms.
The most common approach involves reducing individual payload sizes, typically through data compression \cite{ID_A2, ID_126, ID_A70, ID_A30, ID_47, ID_38}.
Conversely, it can involve aggregating smaller packets to mitigate the protocol overhead induced by handshakes and ACK transmissions \cite{ID_A56}, thereby achieving an energy-efficient transmission threshold \cite{ID_60, ID_21, ID_A7, ID_A44, ID_124}. 
Although the name \textsf{reduce payload} intuitively suggests only size reduction, reflecting its predominant usage in the literature, payload aggregation represents an equally viable counterpart aimed at optimising overall transmission efficiency. 

Finally, rather than focussing on transmission payload size, the \textsf{use dynamic retry} refactoring involves dynamically adjusting the time interval between failed attempts when accessing resources, such as remote databases or network services~\cite{ID_47, ID_38}.

\subsection{Excessive Encapsulation}
\label{sec:classification:EE}
\noindent
The \textsf{Excessive Encapsulation} smell occurs when software strictly meets design-pattern principles that prioritise non-functional requirements -- such as maintainability, modularity, and extensibility -- over energy efficiency \cite{ID_99}.
This smell typically manifests when accessing fields via private or internal getter and setter methods introduces extra function call overhead, leading to runtime performance degradation and increased energy consumption \cite{ID_94, ID_A15}.

Intuitively, the \textsf{reduce encapsulation} refactoring is applied to counteract the negative impacts of this smell. 
To access data without encapsulation overhead, getter and setter method calls are either replaced with direct field access \cite{ID_46, ID_A29, ID_A69, ID_A7, ID_A10, ID_A3} or inlined by adjusting field visibility-access permissions \cite{ID_44}.
This smell is particularly costly in mobile environments such as Android \cite{ID_86}, where the energy-saving impact of this refactoring has been empirically demonstrated \cite{ID_21, ID_A68, ID_A30, ID_A6}. In Android, for instance, even trivial getter or setter invocations are frequently compiled into virtual method calls, incurring non-negligible performance and energy overheads \cite{ID_A31}.

\subsection{Redundant Code}
\label{sec:classification:RC}
\noindent
The \textsf{Redundant Code} smell is due to outdated, duplicated (\aka spaghetti, dead) code that remains within the codebase despite never being executed.
Although inactive, such code must still be loaded into memory, thereby incurring unnecessary energy consumption \cite{ID_A54, ID_48, ID_5, ID_116}. 
This smell often arises from strict adherence to rigorous structures and abstractions aimed at maintainability \cite{ID_21}, or through speculative over-engineering -- such as incorporating unused features or code designed for anticipated future requirements \cite{ID_129}.
Even if dead code does not trigger runtime exceptions directly, it increases software complexity and maintenance overhead, expanding the frequency of software bugs \cite{ID_25, ID_59}.

Intuitively, the \textsf{remove obsolete code} refactoring mitigates the energy impact of this smell \cite{ID_A2, ID_A70, ID_57, ID_100}.
For instance, when identical code fragments are duplicated across different parts of the software, this refactoring replaces duplicate blocks with method invocations or leverages proper object-oriented inheritance \cite{ID_129}.
When duplication occurs within constructors, constructor chaining or delegating calls to the superclass constructor can be applied \cite{ID_59, ID_A35}.
Alternatively, if the smell manifests as redundant memory accesses to the same data structures, unifying data stores and consolidating access patterns eliminates unnecessary overhead \cite{ID_A33, ID_A15, ID_A24}.

\subsection{Excessive Event Rate}
\label{sec:classification:EER}
\noindent
The \textsf{Excessive Event Rate} smell occurs when data or events are continuously processed while the underlying device remains in an idle state or fails to act upon them \cite{ID_A7}.
Energy consumption scales up as either the number of event emitters or their emission rates increase \cite{ID_60}.
Typical manifestations include high-frequency sensor readings and alarm triggers, continuous network polling, frequent database transactions, repetitive update notifications, or iterative queries to code generation models until a stationary output is reached.

The \textsf{reduce event rate} refactoring aims at decreasing overall event frequency, such as, by extending the time interval between consecutive events \cite{ID_A8, ID_A57, ID_60, ID_21, ID_88, ID_A2, ID_49, ID_47}.
For instance, this smell occurs when a monitoring service continuously detects and restarts an application that was explicitly terminated by the user, resulting in a loop bug that always returns the same initial state.
In such case, the refactoring narrows the event rate, such as, trying to poll the server three times before giving up \cite{ID_A24}.

To mitigate the energy impact of this smell, the \textsf{push over pull} refactoring replaces active  resource querying (\ie polling) with push-based notifications \cite{ID_A8, ID_49, ID_47, ID_38}.
Alternatively, the \textsf{batch events} refactoring consolidates multiple operations into batch processing tasks \cite{ID_A8, ID_60, ID_A2, ID_A56, ID_A48, ID_47}.
For instance, when dealing with rigidly scheduled timers or alarms, this refactoring synchronises multiple recurring alerts to execute at specific, unified time intervals \cite{ID_94, ID_A54, ID_49, ID_A30}.

\subsection{Excessive Indirection}
\label{sec:classification:EI}
\noindent
The \textsf{Excessive Indirection} smell focuses on runtime object creation overheads that occur when software strictly adheres to design patterns \cite{ID_46, ID_67} (\eg Observer \cite{ID_21, ID_99, ID_A70}, Abstract Factory \cite{ID_21}, Decorator \cite{ID_21, ID_99}, Mediator \cite{ID_99}) to prioritise non-functional requirements (\eg readability, maintainability) over energy efficiency. 
The negative impacts of this smell stem from complex, cyclic dependencies that can trigger cascading energy-consuming effects \cite{ID_129, ID_130}. For instance, this smell occurs when reducing method calls does yields no energy efficiency gain because individual invocations pass excessive variables, incurring high data transfer overheads during inter-method communications \cite{ID_A28, ID_A12, ID_A24}.

The \textsf{reduce indirection} refactoring is applied to counteract the negative impacts of this smell. 
The implementation either replaces a method call directly with the callee's body (\aka inline method) \cite{ID_99, ID_46, ID_A28, ID_25, ID_A7, ID_A3, ID_67,  ID_A12, ID_A24, ID_100} or reduces communication costs by grouping repeated parameters into a single object and updating callers to pass this instance (\aka introduce parameter object, data clumps) \cite{ID_A70, ID_130, ID_67}. The latter clusters data to diminish dependencies among calling components and reduce the resulting energy consumption \cite{ID_129}.
Alternatively, at a finer granularity level, the refactoring (\aka inline temp), eliminates temporary variables by replacing their references directly with the expression or function call yielding the value \cite{ID_44}.

\subsection{Excessive Delegation}
\label{sec:classification:ED}
\noindent
From an object-oriented perspective, the \textsf{Excessive Delegation} smell occurs when a class is made \enquote{excessively} flexible or modular \cite{ID_A19, ID_129, ID_A3}, often through the strict application of design patterns \cite{ID_126,ID_120} (\eg typical implementations of Visitor \cite{ID_46}, Decorator \cite{ID_21, ID_99, ID_86}, Observer \cite{ID_21, ID_99, ID_A69}, or Mediator \cite{ID_99}). Traversing deep delegation chains in such designs introduces runtime overhead, particularly when invoked frequently, and can trigger cascading events if not properly managed \cite{ID_60}. Furthermore, managing and invoking components that are excessively decoupled, often implemented by others, can be additionally complicated and therefore energy consuming \cite{ID_25, ID_124}.

The \textsf{reduce delegation} refactoring enables mitigating the negative impacts of this smell. The implementation shifts structure definition from runtime delegation toward compile-time inheritance \cite{ID_A19, ID_21, ID_A3, ID_44}. By collapsing data and logic behaviour into the mediating class \cite{ID_129}, this transformation is achieved either through the removal of complex design patterns \cite{ID_60, ID_99, ID_46, ID_A69} or via direct code restructuring.
For instance, macro functions can be used at compile time as an alternative technique to implement this refactoring, thereby erasing function call overhead and jumps to external function bodies \cite{ID_25}.

\subsection{Long Function}
\label{sec:classification:LF}
\noindent
The \textsf{Long Function} smell occurs when a method or function implementation becomes overly long or complex, directly contributing to increased energy consumption \cite{ID_A33, ID_21, ID_46, ID_48, ID_A35, ID_44, ID_A15}. 
This smell is typically detected through quantitative software metrics -- such as lines of code, executable statements, branch counts, variables used, and parameters passed -- or high-level logical indicators, including functional complexity and code coverage levels \cite{ID_A54, ID_129, ID_A10, ID_86, ID_59}.
In software evolution, this smell tends to manifest in later application releases as features accumulation increases overall system complexity over time \cite{ID_A33,ID_A15}.

The \textsf{decompose function} refactoring involves splitting a long method or function into smaller, more manageable units and, if necessary, move them to appropriate classes to improve execution efficiency \cite{ID_A33, ID_48, ID_A35, ID_A15}.
Implementing this refactoring relies on satisfying established quantitative and qualitative software standards, such as reducing total method length \cite{ID_59} or capping parameter lists to a maximum threshold (\eg no more than five parameters) \cite{ID_129}.

However, function decomposition must yield energy reduction without introducing alternative green smells, such as excessive modularisation. 
For instance, refactoring a method with an excessive parameter list alters the caching behaviour
While passing fewer parameters reduces CPU overhead during method invocation, it can increase overall memory transactions. Conversely, encapsulating all parameters within a single method allows initial bulk caching—avoiding repetitive memory loads—at the cost of increased initial CPU load to process parameters that may not be immediately required. Decoupling such methods disrupts this execution pipeline, directly shifting the trade-off between CPU processing cycles and memory bandwidth \cite{ID_129}.

\subsection{Repeated Computation}
\label{sec:classification:RC}
\noindent
The \textsf{Repeated Computation} smell refers to the redundant execution of operations characterised by high intrinsic energy demands -- such as I/O-intensive tasks, non-local memory accesses, or complex expression evaluations. 
Given their resource-intensive nature, the execution frequency of these operations should be minimised.
For instance, when this smell involves the repetitive evaluation of complex expressions, the \textsf{Extract Local Variable} refactoring assigns the expression result to a new temporary variable and replaces subsequent redundant valuations with references to that variable \cite{ID_A70, ID_113, ID_67}.

Alternatively, when this smell involves I/O-intensive operations or frequent non-local memory accesses \cite{ID_48}, applying either the \textsf{use caching} or \textsf{use pre-fetching} refactoring can help mitigate the energy impact of this smell.
Under the \textsf{use caching} refactoring, execution is triggered only when the cached object state changes or when performance constraints are relaxed, thereby eliminating redundant invocations and reducing overall energy consumption \cite{ID_18, ID_A70, ID_A69, ID_A7, ID_A1, ID_47, ID_38}.
On the other hand, the \textsf{use pre-fetching} refactoring proactively executes operations anticipated in the near future -- a typical case is the data structure pre-initialisation \cite{ID_60, ID_A69}.

\subsection{Inefficient Parallelism}
\label{sec:classification:IP}
\noindent
Focused on improper thread management, the \textsf{Inefficient Parallelism} smell increases energy consumption.
Examples of this smell include starting a thread without properly terminating it upon task completion (\aka LeakingThread), acquiring but failing to release a wakelock or using one unnecessarily (\aka WakeLock, DurableWakelock), and introducing excessive parallelisation unsuited to the target execution context \cite{ID_77, ID_94}.
While parallelism aims at maximising hardware resource utilisation through concurrent task execution, unmanaged concurrency can lead to thread leaks.

To address this smell, the \textsf{avoid leaking threads} refactoring ensures efficient execution by preventing orphan or leaking threads from consuming system resources \cite{ID_57}. 
In practice, this refactoring often employs running a check variable to track thread state and safely terminate leaking threads on that variable \cite{ ID_86}. 
Alternatively, the refactoring implementation includes to properly release wake locks when unnecessary.
A typical example of these cases is before an application switches to the background \cite{ID_A27}  or by enforcing a release timeout after which the lock automatically expires \cite{ID_A28, ID_88, ID_A29, ID_A30}.
Finally, the \textsf{limit threads} refactoring dynamically scales the number of active worker threads to establish an optimal balance between concurrency levels and the underlying execution context \cite{ID_48, ID_57}.
This is typically achieved by aligning the active thread count with the number of available logical CPU cores \cite{ID_124}.

\subsection{Wrong Variable Type}
\label{sec:classification:ED}
\noindent
The \textsf{Wrong Variable Type} smell occurs when a variable is declared using an energy-inefficient data type and a more optimal alternative is available, often leading to repeated and costly type checking operations at runtime.
Specifically, conversion complexity frequently arises when data is stored in proprietary or complex formats \cite{ID_A1}, requiring iterative accessing and type verification across different values (\aka TypeChecking) \cite{ID_A33}.

This smell typically occurs when repeatedly accessing global variables within loops, incurring higher operational overhead due to pointer dereferencing toward global memory access compared to stack-allocated memory. 
To mitigate this impact, the \textsf{move global to local variables} refactoring introduces local variables within the loop scope \cite{ ID_110, ID_25,ID_100}.
Alternatively, this smell also occurs when relying on static variables, which empirical evidence indicates consume more energy than instance variables \cite{ID_126, ID_116}. 
To mitigate this impact, the \textsf{move static to instance variables} refactoring converts static storage into instance-level definitions \cite{ID_A33, ID_A35, ID_67, ID_A9, ID_A15}.

\subsection{Inefficient Algorithm}
\label{sec:classification:IA}
\noindent
The \textsf{Inefficient Algorithm} smell stems from an energy-intensive algorithm selection \cite{ID_113, ID_124, ID_100}.
This smell may arise not only from a lack of awareness regarding green coding practices but also from development shortcuts -- such as reimplementing existing functionalities with suboptimal energy performance.
A typical example occurs when developers manually copy arrays instead of leveraging optimised, built-in library functions \cite{ID_21}.

The \textsf{use optimised algorithm} refactoring targets improvements at either implementation or selection level of an algorithm \cite{ID_5}. 
Employing algorithms designed to reduce execution time, minimise memory usage, reuse objects, and follow established design patterns--primarily to satisfy performance requirements-- inherently yields energy-efficiency gains as well \cite{ID_57}.
Alternatively, empirical measurements can directly guide the selection of more energy-efficient algorithm variants \cite{ID_A4}.
Examples including selecting graph cut over simulated annealing for stereo matching \cite{ID_67}, replacing bubble sort with quicksort \cite{ID_A3}, or substituting quicksort with insertion sort for small dataset sizes \cite{ID_A7}.

\subsection{Inefficient third-party APIs}
\label{sec:classification:ITA}
\noindent
The \textsf{Inefficient third-party APIs} smell occurs when the reliance on third-party APIs negatively affects energy consumption \cite{ID_A15}.
A prominent example is the inclusion of third-party advertising pop-ups \cite{ID_21, ID_25}.
To address this issue, the \textsf{use energy-efficient APIs} refactoring aims at mitigating energy impact by reducing or optimising third-party advertising services \cite{ID_88}.

In practice, when this smell occurs from the usage of energy-greedy third-party APIs \cite{ID_126}, the \textsf{use energy-efficient APIs} refactoring mitigates its energy impact by avoiding their usage or opting for energy-efficient alternative APIs \cite{ID_49, ID_47}.
For instance, during file-copy operations, developers should select APIs that offer tunable buffer sizes rather than relying on interfaces restricted to fixed-size buffers \cite{ID_A40}.

\subsection{Single Functioning}
\label{sec:classification:SF}
\noindent
The \textsf{Single Functioning} smell occurs when software exhibits a static energy profile, failing to adapt its feature set based on the energy availability during execution.
A primary manifestation of this smell is the failure to restrict non-essential functionality during idle states or at system boot \cite{ID_49}.  
Furthermore, energy consumption is often influenced by uncontrollable environmental and system-level factors, such as ambient temperature, tail energy consumption, or concurrent background tasks. 
If unmanaged, these factors progressively degrade overall energy efficiency unless the computational load is dynamically adjusted \cite{ID_88}.

The negative effects of this smell can be mitigated by making software power aware. 
Specifically, by refactoring the system to \textsf{enable} an \textsf{emergency mode} that tailors the software with offered features to the current power state. 
Such adaptive behaviour can be implemented through various strategies. 
For instance, when operating on a low battery, a mobile device may retrieve plain text rather than high-fidelity images or video streams \cite{ID_A8, ID_A55}, thereby prioritising core data delivery over non-essential processing \cite{ID_47}. 
Alternatively, non-urgent computations can be deferred to more appropriate execution windows \cite{ID_94}, run under restricted memory constraints \cite{ID_94, ID_A30}, or transitioned to low-power communication protocols where available.
A typical example of the latter case is switching from standard Bluetooth to Bluetooth Low Energy \cite{ID_49}.

\subsection{Interpreted Programming Language}
\label{sec:classification:IPL}
\noindent
The \textsf{Interpreted Programming Language} smell arises when programming language selection -- often driven by company policies or developer preference -- introduces inherent energy overhead \cite{ID_126, ID_75}.  
In practice, optimising programming language selection targets a compilation closer to native machine code, given that execution runtime performance is the primary factor influencing software energy efficiency \cite{ID_46, ID_57}. 

For instance, scientific computing applications often involve intensive computation, causing language choice to significantly impact power usage. 
Additionally, the development context can either amplify or mitigate this influence. 
Consequently, the \textsf{switch to compiled language} refactoring aims at replacing the programming language or execution runtime of a function with a more energy-efficient alternative.
A typical example of allowing the switch toward a different programming language, without introducing not so heavy overheads, is by leveraging on serverless computing platforms \cite{ID_24}.

\subsection{Useless Recursion}
\label{sec:classification:UR}
\noindent
The least frequent smell is the \textsf{Useless Recursion} smell, which occurs when a recursive function implementation unnecessarily elevates energy consumption.
Alternatively, this smell could also occur when recursively calling a function incur high computational overhead to allocate and manage execution states -- including stack frames and local variables -- before, during, and after recursion completes.

Consequently, this smell increases energy usage by accumulating execution resources -- most notably memory allocation across the function call stack -- required to preserve state throughout the overall processing lifecycle \cite{ID_25}. 
To mitigate this, the \textsf{use iterative solutions} refactoring transforms recursive structures into iterative implementations, thereby eliminating call-stack overhead and yielding energy savings \cite{ID_25, ID_113, ID_100}.

\section{Threats to Validity}
\label{sec:ttv}

\noindent
Wohlin et al. \cite{Wohlin2000_ExperimentationSE} define potential threats to the validity of studies in empirical software engineering, four of which may apply to the validity of our study.
These are the threats to \textit{external}, \textit{internal}, \textit{construct}, and \textit{conclusions} validities, which we discuss hereafter.

\smallskip \noindent 
\textbf{External Validity}.
The external validity of a study affect the applicability of the study's results in a more general context \cite{Wohlin2000_ExperimentationSE}.
Since the selected studies considered by our review are obtained from a large extent of online available sources, the identified green smells and corresponding refactorings may be only partly applicable to the broad area of software environmental sustainability. This may potentially threaten the external validity of our study.
To mitigate this potential threat, we organised multiple feedback sessions during our analysis of the selected studies. 
The discussions held within and after the feedback session resulted in qualitative data, which we exploited to fine-tune both our research methods and the applicability of our findings. 
We also prepared a replication package bundling the intermediary artefacts and the final results produced while running our study, which is publicly available online.\footnote{
\url{https://l1nq.com/0ehsfw4}
}
The replication package, together with the references to the selected studies listed in \Cref{tab:design:selected-studies}, can be used to deepen the understanding of our data, as well as to repeat our study and verify our findings.

Additionally, the external validity of our study may be threatened by the risk of having missed relevant studies, since the concepts related to those included in our search strings are differently named in such studies.
To mitigate this potential threat, we have explicitly included all relevant synonyms in our search strings and we also enacted backward snowballing (see \Cref{sec:design}).

\smallskip \noindent 
\textbf{Internal Validity}
The internal validity of a study concerns the validity of the method employed to analyse the study data \cite{Wohlin2000_ExperimentationSE}.
To reinforce the internal validity of our study, we relied on established guidelines to conduct systematic literature reviews in software engineering \cite{Kitchenham2007_GuidelinesSLRinSE,Petersen2008_SMSGuidelines,Petersen2015_SMSGuidelinesUpdate}.
Additionally, we enacted various triangulation rounds and adopted inter-rater reliability assessment (see \Cref{sec:design}), which were conceived to avoid possible biases by triangulation.

\smallskip \noindent 
\textbf{Construct and Conclusions Validity}
The construct and conclusions validities concerns the generalisability of the constructs 
and the degree to which the conclusions of a study are reasonably based on the available data, respectively \cite{Wohlin2000_ExperimentationSE}.
The construct and conclusions validities of our study may have been affected by observer and interpretation biases.
To mitigate these potential threats, we exploited inter-rater reliability assessment (see \Cref{sec:design}), and we also independently drawn conclusions from the available data, which were then discussed to reach a shared consensus on the green smells and refactorings presented in this article.

\section{Related Work}
\label{sec:related}

\noindent
Various secondary studies analyse and classify the state-of-the-art approaches for environmentally sustainable software.
A first example is given by Georgiou et al. \cite{Georgiou_SDL_energy_efficiency}, who survey all software development lifecycle phases for discovering approaches, tools, and techniques for energy-efficient programming. 
Kosbar and Hamdaqa \cite{kosbar2025smellssussustainabilitysmellsiac} review existing solutions for sustainable software deployment over IT infrastructures.
The above listed studies differ from ours in their objective, as they discuss how to achieve environmentally sustainable software from a different perspective than that of green smells and refactorings. 
We instead aim at providing a first catalogue of green smells, namely of the bad decisions in design or development that could possibly affect the environmental sustainability of software, and of the refactorings enabling to mitigate their effects.

Similar arguments apply to Durdik et al. \cite{Durdik_guidelines} and to our previous work \cite{Danushi2025_SLRCarbonEfficientSW}. Durdik et al. \cite{Durdik_guidelines} review the existing guidelines for achieving sustainability in long-lived systems, however focusing on financial (rather than environmental) sustainability. Instead, our previous work \cite{Danushi2025_SLRCarbonEfficientSW} recaps the state-of-the-art solutions for designing and developing carbon-efficient software.
Differently from Durdik et al. \cite{Durdik_guidelines} and our previous work \cite{Danushi2025_SLRCarbonEfficientSW}, we here elicit a first catalogue of green smells and their corresponding refactorings.

Other noteworthy secondary studies are the literature reviews by Xiao et al. \cite{Xiao_2025} and Albonico et al. \cite{Albonico_robots}, who investigate energy-efficient patterns and evaluate the trade-offs between reducing energy consumption and satisfying other software quality requirements. 
Similarly, Gaglianese et al.~\cite{Gaglianese2023_GreenCloudEdgeOrchestration} survey the existing approaches for the environmentally sustainable orchestration of containerised software applications over Cloud-Edge infrastructures.
These studies focus on formulating energy-efficiency guidelines to design, develop, or deploy software systems.
We instead focus on deriving a catalogue of green smells and refactorings.

The work most closely related to ours is by Le Goaër \cite{ID_49}, which is included among the primary studies considered by our SLR.
Le Goaër \cite{ID_49} first proposed a catalogue of green smells and refactorings for Android mobile software. We here expand the catalogue of green smells and refactorings to cover a broader scope, independently from the application domain, and by providing an up-to-date analysis ofl the relevant literature on the topic.

In summary, to the best of our knowledge, our SLR is the first providing a catalogue of green smells denoting bad decisions in software design or development that could possibly affect its environmental sustainability, and of the refactorings enabling to mitigate their effects.

\section{Conclusions}
\label{sec:conclusions}

\noindent
We presented the results of an SLR focused on identifying the smells denoting possible software energy consumption, as well as the refactorings enabling to mitigate their negative impacts. 
Specifically, we presented a taxonomy organising 20 green smells and 33 refactorings, provided a detailed analysis of when each smell occurs, and we showed how corresponding refactorings effectively address them.

The results of our study can be of help to both researchers and practitioners interested in designing and developing environmentally sustainable software. 
Our study indeed complements the existing study on Android mobile development by Le Goaër \cite{ID_49} by providing a domain-agnostic software engineering perspective.
From a pragmatic standpoint, software architects and developers can leverage our results in their daily work make their software more sustainable from an environmental perspective. In this perspective, we plan to devise a decision support, \eg based on ATAM trees or decision models, which would enable architects/developers to consider whether to refactor a green smell, not only considering its environmental impact but also the impacts that such refactoring would have on other software quality attributes, \eg maintainability or performance efficiency.

For future work, we also plan to study how to automatically identify green smells, as well as how to automatically or semi-automatically implement their corresponding refactorings. 
Our ultimate goal would indeed be to provide software architects and developers with a full-fledged support for resolving the green smells in their software.

\bibliographystyle{elsarticle-harv}
\bibliography{biblio.bib}

@article{Danushi2025_SLRCarbonEfficientSW,
  author = {Danushi, Ornela and others},
  title = {Carbon-Efficient Software Design and Development: A Systematic Literature Review},
  year = {2025},
  issue_date = {October 2025},
  publisher = {Association for Computing Machinery},
  address = {New York, NY, USA},
  volume = {57},
  number = {10},
  issn = {0360-0300},
  doi = {10.1145/3728638},
  journal = {ACM Comput. Surv.},
  month = may,
  articleno = {267},
  numpages = {35}}

@article{Jukka,
    author = {Manner, Jukka},
    title = {Black software — the energy unsustainability of software systems in the 21st century},
    journal = {Oxford Open Energy},
    volume = {2},
    pages = {oiac011},
    year = {2023},
    month = {01},
    issn = {2752-5082},
    doi = {10.1093/ooenergy/oiac011}}

@InProceedings{Fowler_refactoring,
author="Fowler, Martin",
editor="Wells, Don
and Williams, Laurie",
title="Refactoring: Improving the Design of Existing Code",
booktitle="Extreme Programming and Agile Methods --- XP/Agile Universe 2002",
year="2002",
publisher="Springer Berlin Heidelberg",
address="Berlin, Heidelberg",
pages="256--256",
isbn="978-3-540-45672-8"}

@inproceedings{Petersen2008_SMSGuidelines, 
    author = {Petersen, Kai and Feldt, Robert and Mujtaba, Shahid and Mattsson, Michael}, 
    title = {Systematic mapping studies in software engineering}, 
    year = {2008}, 
    publisher = {BCS Learning and Development Ltd.}, 
    address = {Swindon, GBR}, 
    booktitle = {Proceedings of the 12th International Conference on Evaluation and Assessment in Software Engineering}, 
    pages = {68–77}, 
    numpages = {10}, 
    location = {Italy}, 
    series = {EASE'08}}

@article{Petersen2015_SMSGuidelinesUpdate,
    title = {Guidelines for conducting systematic mapping studies in software engineering: An update},
    journal = {Information and Software Technology},
    volume = {64},
    pages = {1-18},
    year = {2015},
    issn = {0950-5849},
    doi = {10.1016/j.infsof.2015.03.007},
    author = {Kai Petersen and Sairam Vakkalanka and Ludwik Kuzniarz}}

@misc{Kitchenham2007_GuidelinesSLRinSE,
  author = {Kitchenham, Barbara Ann and Charters, Stuart},
  title = {Guidelines for performing Systematic Literature Reviews in Software Engineering},
  howpublished = {\textit{Keele University and Durham University Joint Report}, EBSE 2007-001},
  year = {2007}}

@book{Krippendorff2004_ContentAnalysis,
    author = {Klaus Krippendorff},
    title = {Content Analysis: An Introduction to Its Methodology (2 ed.).},
    publisher = {Sage, Thousand Oaks, CA.},
    year = {2004}}

@inproceedings{Wohlin2014_GuidelinesSnowballing,
  author       = {Claes Wohlin},
  editor       = {Martin J. Shepperd and others},
  title        = {Guidelines for snowballing in systematic literature studies and a
                  replication in software engineering},
  booktitle    = {18th International Conference on Evaluation and Assessment in Software
                  Engineering, {EASE} '14, London, England, United Kingdom, May 13-14,
                  2014},
  pages        = {38:1--38:10},
  publisher    = {{ACM}},
  year         = {2014},
  doi          = {10.1145/2601248.2601268}}

@INPROCEEDINGS{ID_3,
  author={Cordero, Víctor and Guzmán, Ignacio García-Rodríguez de and Piattini, Mario},
  booktitle={IEEE 10th International Conference on Global Software Engineering Workshops}, 
  title={A First Approach on Legacy System Energy Consumption Measurement}, 
  year={2015},
  volume={},
  number={},
  pages={35-43},
  doi={10.1109/ICGSEW.2015.15}}

@INPROCEEDINGS{ID_5,
  author={O'Dea, Eoan},
  booktitle={IEEE 22nd International Conference on Software Architecture Companion (ICSA-C)}, 
  title={A Measurement-Driven Approach to Enhancing Sustainability in Microservice Architectures}, 
  year={2025},
  volume={},
  number={},
  pages={167-172},
  doi={10.1109/ICSA-C65153.2025.00031}}

@INPROCEEDINGS{ID_10,
  author={Dhaka, Garima and Singh, Paramvir},
  booktitle={23rd Asia-Pacific Software Engineering Conference (APSEC)}, 
  title={An Empirical Investigation into Code Smell Elimination Sequences for Energy Efficient Software}, 
  year={2016},
  volume={},
  number={},
  pages={349-352},
  doi={10.1109/APSEC.2016.057}}

@INPROCEEDINGS{ID_18,
  author={Giavrimis, Rafail and Basios, Michail and Wu, Fan and Kanthan, Leslie and Bauer, Roman},
  booktitle={IEEE Conference on Artificial Intelligence (CAI)}, 
  title={Artemis AI: Multi-LLM Framework for Code Optimisation}, 
  year={2025},
  volume={},
  number={},
  pages={1-6},
  doi={10.1109/CAI64502.2025.00057}}

@INPROCEEDINGS{ID_21,
  author={Connolly Bree, Deaglan and Cinneide, Mel O},
  booktitle={IEEE International Conference on Software Maintenance and Evolution (ICSME)}, 
  title={Automated Refactoring for Energy-Aware Software}, 
  year={2021},
  volume={},
  number={},
  pages={689-694},
  doi={10.1109/ICSME52107.2021.00082}}

@INPROCEEDINGS{ID_24,
  author={Werner, Sebastian and Kähler, Mathis and Hakamian, Alireza},
  booktitle={IEEE International Conference on Cloud Engineering (IC2E)}, 
  title={Code once, Run Green: Automated Green Code Translation in Serverless Computing}, 
  year={2025},
  volume={},
  number={},
  pages={105-113},
  doi={10.1109/IC2E65552.2025.00026}}

@article{ID_25,
  title={Code refactoring techniques for reducing energy consumption in embedded computing environment},
  author={Doohwan Kim and Jang-Eui Hong and Il-Cheol Yoon and Sang-Ho Lee},
  journal={Cluster Computing},
  year={2016},
  volume={21},
  pages={1079 - 1095}}

@INPROCEEDINGS{ID_38,
  author={Ribeiro, Ana and Ferreira, João F. and Mendes, Alexandra},
  booktitle={IEEE 21st International Conference on Software Quality, Reliability and Security (QRS)}, 
  title={EcoAndroid: An Android Studio Plugin for Developing Energy-Efficient Java Mobile Applications}, 
  year={2021},
  volume={},
  number={},
  pages={62-69},
  doi={10.1109/QRS54544.2021.00017}}

@article{ID_44,
author = {Sanlıalp, Ibrahim and Ozturk, Muhammed and Yigit, Tuncay},
year = {2022},
month = {02},
pages = {442},
title = {Energy Efficiency Analysis of Code Refactoring Techniques for Green and Sustainable Software in Portable Devices},
volume = {11},
journal = {Electronics},
doi = {10.3390/electronics11030442}}

@article{ID_46,
author = {Connolly Bree, Deaglan and O Cinneide, Mel},
title = {Energy efficiency of the Visitor Pattern: contrasting Java and C++ implementations},
year = {2023},
publisher = {Kluwer Academic Publishers},
address = {USA},
volume = {28},
number = {6},
issn = {1382-3256},
doi = {10.1007/s10664-023-10387-8},
numpages = {36}}

@inproceedings{ID_47,
author = {Rani, Pooja and Zellweger, Jonas and Kousadianos, Veronika and Cruz, Luis and Kehrer, Timo and Bacchelli, Alberto},
title = {Energy Patterns for Web: An Exploratory Study},
year = {2024},
isbn = {9798400704994},
publisher = {Association for Computing Machinery},
address = {New York, NY, USA},
doi = {10.1145/3639475.3640110},
booktitle = {Proceedings of the 46th International Conference on Software Engineering: Software Engineering in Society},
pages = {12–22},
numpages = {11},
location = {Lisbon, Portugal},
series = {ICSE-SEIS'24}}

@INPROCEEDINGS{ID_48,
  author={Jolly, Sonali and Yadav, Asmita and Singh, Sandeep Kumar},
  booktitle={17th International Conference on Contemporary Computing (IC3)}, 
  title={Energy-Aware Computing: Cutting CPU Cycles through Intelligent Thread Management and Code Smell Optimization}, 
  year={2025},
  volume={},
  number={},
  pages={1-7},
  doi={10.1109/IC366947.2025.11290127}}

@INPROCEEDINGS{ID_49,
  author={Goaër, Olivier Le},
  booktitle={35th IEEE/ACM International Conference on Automated Software Engineering Workshops (ASEW)}, 
  title={Enforcing Green Code With Android Lint}, 
  year={2020},
  volume={},
  number={},
  pages={85-90},
  doi={10.1145/3417113.3422188}}

@INPROCEEDINGS{ID_57,
  author={Toosi, Farshad Ghassemi},
  booktitle={Computing, Communications and IoT Applications (ComComAp)}, 
  title={Green Software Engineering: A Dual-Perspective Overview of Stakeholder and Societal Interpretations}, 
  year={2025},
  volume={},
  number={},
  pages={385-394},
  doi={10.1109/ComComAp68359.2025.11353190}}

@article{ID_59,
author = {Sehgal, Rajni and Mehrotra, Deepti and Nagpal, Renuka and Sharma, Ramanuj},
title = {Green software: Refactoring approach},
year = {2022},
publisher = {Elsevier Science Inc.},
address = {USA},
volume = {34},
number = {7},
issn = {1319-1578},
doi = {10.1016/j.jksuci.2020.10.022},
journal = {J. King Saud Univ. Comput. Inf. Sci.},
month = jul,
pages = {4635–4643},
numpages = {9}}

@INPROCEEDINGS{ID_60,
  author={Chowdhury, Shaiful Alam and Hindle, Abram and Kazman, Rick and Shuto, Takumi and Matsui, Ken and Kamei, Yasutaka},
  booktitle={IEEE/ACM 41st International Conference on Software Engineering (ICSE)}, 
  title={GreenBundle: An Empirical Study on the Energy Impact of Bundled Processing}, 
  year={2019},
  volume={},
  number={},
  pages={1107-1118},
  doi={10.1109/ICSE.2019.00114}}

@inproceedings{ID_67,
author = {Sahin, Cagri and Pollock, Lori and Clause, James},
title = {How do code refactorings affect energy usage?},
year = {2014},
isbn = {9781450327749},
publisher = {Association for Computing Machinery},
address = {New York, NY, USA},
doi = {10.1145/2652524.2652538},
booktitle = {Proceedings of the 8th ACM/IEEE International Symposium on Empirical Software Engineering and Measurement},
articleno = {36},
numpages = {10},
location = {Torino, Italy},
series = {ESEM '14}}

@INPROCEEDINGS{ID_75,
  author={Pereira, Rui and Simão, Pedro and Cunha, Jácome and Saraiva, João},
  booktitle={2018 33rd IEEE/ACM International Conference on Automated Software Engineering (ASE)}, 
  title={jStanley: Placing a Green Thumb on Java Collections}, 
  year={2018},
  volume={},
  number={},
  pages={856-859},
  doi={10.1145/3238147.3240473}}

@INPROCEEDINGS{ID_77,
  author={Cruz, Luis and Abreu, Rui and Rouvignac, Jean-Noël},
  booktitle={IEEE/ACM 4th International Conference on Mobile Software Engineering and Systems (MOBILESoft)}, 
  title={Leafactor: Improving Energy Efficiency of Android Apps via Automatic Refactoring}, 
  year={2017},
  volume={},
  number={},
  pages={205-206},
  doi={10.1109/MOBILESoft.2017.21}}

@article{ID_86,
title = "On the Impact of Code Smells on the Energy Consumption of Mobile Applications",
author = "Fabio Palomba and \{Di Nucci\}, Dario and Annibale Panichella and Andy Zaidman and \{De Lucia\}, Andrea",
note = "Green Open Access added to TU Delft Institutional Repository {\textquoteleft}You share, we take care!{\textquoteright} – Taverne project https://www.openaccess.nl/en/you-share-we-take-care Otherwise as indicated in the copyright section: the publisher is the copyright holder of this work and the author uses the Dutch legislation to make this work public. ",
year = "2019",
doi = "10.1016/j.infsof.2018.08.004",
volume = "105",
pages = "43--55",
journal = "Information and Software Technology",
issn = "0950-5849",
publisher = "Elsevier"}

@INPROCEEDINGS{ID_88,
  author={Cruz, Luis and Abreu, Rui},
  booktitle={IEEE/ACM 4th International Conference on Mobile Software Engineering and Systems (MOBILESoft)}, 
  title={Performance-Based Guidelines for Energy Efficient Mobile Applications}, 
  year={2017},
  volume={},
  number={},
  pages={46-57},
  doi={10.1109/MOBILESoft.2017.19}}

@ARTICLE{ID_94,
  author={Fawad, Muhammad and Rasool, Ghulam and Bilal Riaz, Muhammad},
  journal={IEEE Access}, 
  title={Refactoring Android Source Code Smells From Android Applications}, 
  year={2025},
  volume={13},
  number={},
  pages={14122-14150},
  doi={10.1109/ACCESS.2025.3529687}}

@INPROCEEDINGS{ID_99,
  author={Connolly Bree, Deaglan and Cinneide, Mel O},
  booktitle={IEEE International Conference on Software Analysis, Evolution and Reengineering (SANER)}, 
  title={Removing Decorator to Improve Energy Efficiency}, 
  year={2022},
  volume={},
  number={},
  pages={902-912},
  doi={10.1109/SANER53432.2022.00108}}

@inproceedings{ID_100,
author = {Yadav, Asmita and Usman, Mufti and Sati, Anurag and Jain, Sidham},
title = {Revolutionizing Software Development: Enhancing Quality and Performance Through Code Refactoring},
year = {2024},
isbn = {9798400709722},
publisher = {Association for Computing Machinery},
address = {New York, NY, USA},
doi = {10.1145/3675888.3676139},
booktitle = {Proceedings of the 2024 Sixteenth International Conference on Contemporary Computing},
pages = {715–725},
numpages = {11},
location = {Noida, India},
series = {IC3-2024}}

@INPROCEEDINGS{ID_110,
  author={Gurung, Ram Prasad and Porras, Jari and Koistinaho, Jarkko},
  booktitle={10th International Conference on ICT for Sustainability (ICT4S)}, 
  title={Static Code Analysis for Reducing Energy Code Smells in Different Loop Types: A Case Study in Java}, 
  year={2024},
  volume={},
  number={},
  pages={292-302},
  doi={10.1109/ICT4S64576.2024.00037}}

@INPROCEEDINGS{ID_113,
  author={Lano, Kevin and Rahimi, Shekoufeh Kolahdouz and Xue, Qiaomu},
  booktitle={ACM/IEEE 28th International Conference on Model Driven Engineering Languages and Systems Companion (MODELS-C)}, 
  title={Sustainability Implications of Large Language Model use in Model-Driven Engineering}, 
  year={2025},
  volume={},
  number={},
  pages={772-776},
  doi={10.1109/MODELS-C68889.2025.00101}}

@inproceedings{ID_116,
author = {Ournani, Zakaria and Rouvoy, Romain and Rust, Pierre and Penhoat, Joel},
year = {2021},
month = {01},
pages = {34-46},
title = {Tales from the Code \#1: The Effective Impact of Code Refactorings on Software Energy Consumption},
doi = {10.5220/0010517900002992}
}

@INPROCEEDINGS{ID_120,
  author={Pereira, Rui and Couto, Marco and Cunha, Jácome and Fernandes, João Paulo and Saraiva, João},
  booktitle={IEEE/ACM 5th International Workshop on Green and Sustainable Software (GREENS)}, 
  title={The Influence of the Java Collection Framework on Overall Energy Consumption}, 
  year={2016},
  volume={},
  number={},
  pages={15-21},
  doi={10.1145/2896967.2896968}}

@inproceedings{ID_124,
author = {Nagy, Gergely and M{\'e}sz{\'a}ros, {\'A}ron Attila and Boz{\'o}, Istv{\'a}n and T{\'o}th, Melinda},
title = {Tools supporting green computing in Erlang},
year = {2019},
isbn = {9781450368100},
publisher = {Association for Computing Machinery},
address = {New York, NY, USA},
doi = {10.1145/3331542.3342570},
booktitle = {Proceedings of the 18th ACM SIGPLAN International Workshop on Erlang},
pages = {30–35},
numpages = {6},
location = {Berlin, Germany},
series = {Erlang 2019}}

@INPROCEEDINGS{ID_126,
  author={Hecktor, Marco and Sachweh, Sabine},
  booktitle={IEEE European Technology and Engineering Management Summit (E-TEMS)}, 
  title={Towards a Conceptual Framework for Greener Software Systems}, 
  year={2025},
  volume={},
  number={},
  pages={62-67},
  doi={10.1109/E-TEMS64751.2025.11239151}}

@INPROCEEDINGS{ID_129,
  author={Imran, Asif and Kosar, Tevfik and Zola, Jaroslaw and Bulut, M. Fatih},
  booktitle={IEEE 17th International Conference on Cloud Computing (CLOUD)}, 
  title={Towards Sustainable Cloud Software Systems through Energy-Aware Code Smell Refactoring}, 
  year={2024},
  volume={},
  number={},
  pages={223-234},
  doi={10.1109/CLOUD62652.2024.00034}}

@inproceedings{ID_130,
  title={Towards the Impact of Design Flaws on the Resources Used by an Application},
  author={Cristina Marinescu and Serban Stoenescu and Teodor-Florin Fortiș},
  booktitle={ARMS-CC@PODC},
  year={2014}}

@article{ID_A1,
author = {Naumann, Stefan and Dick, Markus and Kern, Eva and Johann, Timo},
year = {2011},
month = {12},
pages = {},
title = {The GREENSOFT Model: A reference model for green and sustainable software and its engineering},
volume = {1},
journal = {Surface Science - SURFACE SCI},
doi = {10.1016/j.suscom.2011.06.004}}

@misc{ID_A2,
      title={Innovating for Tomorrow: The Convergence of SE and Green AI}, 
      author={Luís Cruz and Xavier Franch Gutierrez and Silverio Martínez-Fernández},
      year={2024},
      eprint={2406.18142},
      archivePrefix={arXiv},
      primaryClass={cs.SE},
}

@article{ID_A3,
author = {Park, J.-J and Hong, Jang-Eui and Lee, S.-H},
year = {2014},
month = {01},
pages = {717-722},
title = {Investigation for software power consumption of code refactoring techniques},
journal = {Proceedings of the International Conference on Software Engineering and Knowledge Engineering, SEKE}
}

@inproceedings{ID_A4,
author = {Hasan, Samir and King, Zachary and Hafiz, Munawar and Sayagh, Mohammed and Adams, Bram and Hindle, Abram},
title = {Energy profiles of Java collections classes},
year = {2016},
isbn = {9781450339001},
publisher = {Association for Computing Machinery},
address = {New York, NY, USA},
doi = {10.1145/2884781.2884869},
booktitle = {Proceedings of the 38th International Conference on Software Engineering},
pages = {225–236},
numpages = {12},
location = {Austin, Texas},
series = {ICSE '16}}

@inproceedings{ID_A6,
author = {Tonini, Aline and Fischer, Leonardo and Mattos, Julio and Brisolara, Lisane},
year = {2013},
month = {11},
pages = {},
title = {Analysis and Evaluation of the Android Best Practices Impact on the Efficiency of Mobile Applications},
doi = {10.1109/SBESC.2013.39}}

@inproceedings{ID_A7,
author = {Li, Ding and Halfond, William G. J.},
title = {An investigation into energy-saving programming practices for Android smartphone app development},
year = {2014},
isbn = {9781450328449},
publisher = {Association for Computing Machinery},
address = {New York, NY, USA},
doi = {10.1145/2593743.2593750},
booktitle = {Proceedings of the 3rd International Workshop on Green and Sustainable Software},
pages = {46–53},
numpages = {8},
location = {Hyderabad, India},
series = {GREENS 2014}}

@article{ID_A8,
author = {Ayala, Inmaculada and Amor, Mercedes and Fuentes, Lidia},
title = {An Energy Efficiency Study of Web-Based Communication in Android Phones},
journal = {Scientific Programming},
volume = {2019},
number = {1},
pages = {8235458},
doi = {https://doi.org/10.1155/2019/8235458},
year = {2019}}

@article{ID_A9,
  title = {From benchmarks to real apps: Exploring the energy impacts of performance-directed changes},
  author = {Cagri Sahin and Lori L. Pollock and James Clause},
  year = {2016},
  doi = {10.1016/j.jss.2016.03.031},
  journal = {Journal of Systems and Software},
  volume = {117},
  pages = {307-316}}

@article{ID_A10,
author = {Morales, Rodrigo and Saborido Infantes, Rubén and Khomh, Foutse and Chicano, Francisco and Antoniol, Giuliano},
year = {2018},
month = {12},
pages = {1176-1206},
title = {EARMO: An Energy-Aware Refactoring Approach for Mobile Apps},
volume = {44},
journal = {IEEE Transactions on Software Engineering},
doi = {10.1109/TSE.2017.2757486}}

@inproceedings{ID_A12,
author = {Silva, W and Brisolara, Lisane and Corrêa, Ulisses and Carro, Luigi},
year = {2010},
month = {01},
pages = {145-150},
title = {Evaluation of the impact of code refactoring on embedded software efficiency},
doi = {10.13140/2.1.1481.8249}}

@inproceedings{ID_A15,
author = {Verdecchia, Roberto and Saez, René and Procaccianti, Giuseppe and Lago, Patricia},
year = {2018},
month = {02},
pages = {},
title = {Empirical Evaluation of the Energy Impact of Refactoring Code Smells},
doi = {10.29007/dz83}}

@inproceedings{ID_A19,
author = {Connolly Bree, D{\'e}agl{\'a}n and Cinn{\'e}ide, Mel {\'O}},
title = {Inheritance versus Delegation: which is more energy efficient?},
year = {2020},
isbn = {9781450379632},
publisher = {Association for Computing Machinery},
address = {New York, NY, USA},
doi = {10.1145/3387940.3392192},
booktitle = {Proceedings of the IEEE/ACM 42nd International Conference on Software Engineering Workshops},
pages = {323–329},
numpages = {7},
location = {Seoul, Republic of Korea},
series = {ICSEW'20}}

@inproceedings{ID_A24,
author = {Wertjanz, Marion and Josefiok, Mirco and Jelschen, Jan and Winter, Andreas},
year = {2012},
month = {01},
pages = {441-455},
title = {Removing Energy Code Smells with Reengineering Services}}

@misc{ID_A27,
      title={Using Automatic Refactoring to Improve Energy Efficiency of Android Apps}, 
      author={Luis Cruz and Rui Abreu},
      year={2018},
      eprint={1803.05889},
      archivePrefix={arXiv},
      primaryClass={cs.SE}}

@INPROCEEDINGS{ID_A28,
  author={Couto, Marco and Saraiva, João and Fernandes, João Paulo},
  booktitle={IEEE 27th International Conference on Software Analysis, Evolution and Reengineering (SANER)}, 
  title={Energy Refactorings for Android in the Large and in the Wild}, 
  year={2020},
  volume={},
  number={},
  pages={217-228},
  doi={10.1109/SANER48275.2020.9054858}}

@INPROCEEDINGS{ID_A29,
  author={Iannone, Emanuele and Pecorelli, Fabiano and Di Nucci, Dario and Palomba, Fabio and De Lucia, Andrea},
  booktitle={IEEE/ACM 28th International Conference on Program Comprehension (ICPC)}, 
  title={Refactoring Android-specific Energy Smells: A Plugin for Android Studio}, 
  year={2020},
  volume={},
  number={},
  pages={451-455},
  doi={10.1145/3387904.3389298}}

@INPROCEEDINGS{ID_A30,
  author={Palomba, Fabio and Di Nucci, Dario and Panichella, Annibale and Zaidman, Andy and De Lucia, Andrea},
  booktitle={IEEE 24th International Conference on Software Analysis, Evolution and Reengineering (SANER)}, 
  title={Lightweight detection of Android-specific code smells: The aDoctor project}, 
  year={2017},
  volume={},
  number={},
  pages={487-491},
  doi={10.1109/SANER.2017.7884659}}

@INPROCEEDINGS{ID_A31,
  author={Carette, Antonin and Younes, Mehdi Adel Ait and Hecht, Geoffrey and Moha, Naouel and Rouvoy, Romain},
  booktitle={IEEE 24th International Conference on Software Analysis, Evolution and Reengineering (SANER)}, 
  title={Investigating the energy impact of Android smells}, 
  year={2017},
  volume={},
  number={},
  pages={115-126},
  doi={10.1109/SANER.2017.7884614}}

@INPROCEEDINGS{ID_A33,
  author={Anwar, Hina and Pfahl, Dietmar and Srirama, Satish N.},
  booktitle={45th Euromicro Conference on Software Engineering and Advanced Applications (SEAA)}, 
  title={Evaluating the Impact of Code Smell Refactoring on the Energy Consumption of Android Applications}, 
  year={2019},
  volume={},
  number={},
  pages={82-86},
  doi={10.1109/SEAA.2019.00021}}

@InProceedings{ID_A35,
author=" Reeshti and Sehgal, Rajni and Mehrotra, Deepti and Nagpal, Renuka and Choudhury, Tanupriya",
title="Code Smell Refactoring for Energy Optimization of Android Apps",
booktitle="Innovations in Cyber Physical Systems",
year="2021",
publisher="Springer Singapore",
address="Singapore",
pages="371--379",
isbn="978-981-16-4149-7"}

@article{ID_A40,
title = {Impact of Developer Choices on Energy Consumption of Software on Servers},
journal = {Procedia Computer Science},
volume = {62},
pages = {385-394},
year = {2015},
note = {Proceedings of the 2015 International Conference on Soft Computing and Software Engineering (SCSE'15)},
issn = {1877-0509},
doi = {https://doi.org/10.1016/j.procs.2015.08.423},
url = {https://www.sciencedirect.com/science/article/pii/S1877050915025582},
author = {Jasmeet Singh and Kshirasagar Naik and Veluppillai Mahinthan}}

@INPROCEEDINGS{ID_A42,
  author={Agolli, Tedis and Pollock, Lori and Clause, James},
  booktitle={IEEE/ACM 4th International Conference on Mobile Software Engineering and Systems (MOBILESoft)}, 
  title={Investigating Decreasing Energy Usage in Mobile Apps via Indistinguishable Color Changes}, 
  year={2017},
  volume={},
  number={},
  pages={30-34},
  doi={10.1109/MOBILESoft.2017.17}}

@inproceedings{ID_A44,
author = {Li, Ding and Lyu, Yingjun and Gui, Jiaping and Halfond, William G. J.},
title = {Automated energy optimization of HTTP requests for mobile applications},
year = {2016},
isbn = {9781450339001},
publisher = {Association for Computing Machinery},
address = {New York, NY, USA},
doi = {10.1145/2884781.2884867},
booktitle = {Proceedings of the 38th International Conference on Software Engineering},
pages = {249–260},
numpages = {12},
location = {Austin, Texas},
series = {ICSE '16}}

@inproceedings{ID_A46,
author = {Li, Ding and Tran, Angelica Huyen and Halfond, William G. J.},
title = {Making web applications more energy efficient for OLED smartphones},
year = {2014},
isbn = {9781450327565},
publisher = {Association for Computing Machinery},
address = {New York, NY, USA},
doi = {10.1145/2568225.2568321},
booktitle = {Proceedings of the 36th International Conference on Software Engineering},
pages = {527–538},
numpages = {12},
location = {Hyderabad, India},
series = {ICSE 2014}}

@INPROCEEDINGS{ID_A47,
  author={Alam, Faisal and Panda, Preeti Ranjan and Tripathi, Nikhil and Sharma, Namita and Narayan, Sanjiv},
  booktitle={2014 Design, Automation and Test in Europe Conference and Exhibition (DATE)}, 
  title={Energy optimization in Android applications through wakelock placement}, 
  year={2014},
  volume={},
  number={},
  pages={1-4},
  doi={10.7873/DATE.2014.101}}

@inproceedings{ID_A48,
author = {Lyu, Yingjun and Li, Ding and Halfond, William G. J.},
title = {Remove RATs from your code: automated optimization of resource inefficient database writes for mobile applications},
year = {2018},
isbn = {9781450356992},
publisher = {Association for Computing Machinery},
address = {New York, NY, USA},
doi = {10.1145/3213846.3213865},
booktitle = {Proceedings of the 27th ACM SIGSOFT International Symposium on Software Testing and Analysis},
pages = {310–321},
numpages = {12},
location = {Amsterdam, Netherlands},
series = {ISSTA 2018}}

@article{ID_A53,
    author = {Saborido, Ruben and Morales, Rodrigo and Khomh, Foutse and Gueheneuc, Yann-Gael and Antoniol, Giuliano},
    title = {Getting the most from map data structures in android},
    journal = {Empirical Software Engineering, 23(5):2829–2864},
    year = {2018},
    doi = {10.1007/s10664-018-9607-8},
    isbn = {1573-7616}}

@inproceedings{ID_A54,
author = {Gattal, Abderraouf and Hammache, Abir and Bousbia, Nabila and Henniche, Adel Nassim},
title = {Exploiting the progress of OO refactoring tools with Android code smells: RAndroid, a plugin for Android studio},
year = {2021},
isbn = {9781450381048},
publisher = {Association for Computing Machinery},
address = {New York, NY, USA},
doi = {10.1145/3412841.3442129},
booktitle = {Proceedings of the 36th Annual ACM Symposium on Applied Computing},
pages = {1580–1583},
numpages = {4},
location = {Virtual Event, Republic of Korea},
series = {SAC '21}}

@inproceedings{ID_A55,
  title={Energy types},
  author={Michael Cohen and Haitao Steve Zhu and Senem Ezgi Emgin and Yu David Liu},
  booktitle={Conference on Object-Oriented Programming Systems, Languages, and Applications},
  year={2012}}

@inproceedings{ID_A56,
author = {Li, Ding and Halfond, William G. J.},
title = {Optimizing energy of HTTP requests in Android applications},
year = {2015},
isbn = {9781450338158},
publisher = {Association for Computing Machinery},
address = {New York, NY, USA},
doi = {10.1145/2804345.2804351},
booktitle = {Proceedings of the 3rd International Workshop on Software Development Lifecycle for Mobile},
pages = {25–28},
numpages = {4},
location = {Bergamo, Italy},
series = {DeMobile 2015}}

@inproceedings{ID_A57,
author = {Banerjee, Abhijeet and Roychoudhury, Abhik},
title = {Automated re-factoring of Android apps to enhance energy-efficiency},
year = {2016},
isbn = {9781450341783},
publisher = {Association for Computing Machinery},
address = {New York, NY, USA},
doi = {10.1145/2897073.2897086},
booktitle = {Proceedings of the International Conference on Mobile Software Engineering and Systems},
pages = {139–150},
numpages = {12},
location = {Austin, Texas},
series = {MOBILESoft '16}}

@inproceedings{ID_A68,
author = {Moreira, Emanuel and Correia, Filipe F. and Bispo, Jo{\~a}o},
title = {Overviewing the liveness of refactoring for energy efficiency},
year = {2020},
isbn = {9781450375078},
publisher = {Association for Computing Machinery},
address = {New York, NY, USA},
doi = {10.1145/3397537.3397538},
booktitle = {Companion Proceedings of the 4th International Conference on Art, Science, and Engineering of Programming},
pages = {211–212},
numpages = {2},
location = {Porto, Portugal},
series = {Programming '20}}

@article{ID_A69,
    author = {Lano, Kevin and Alwakeel, Lyan and Rahman, Zishan},
    title = {Design Patterns for Software Sustainability},
    journal = {Pattern Languages of Programs},
    year = {2025}
}

@inbook{ID_A70,
title = "Software modelling for sustainable software engineering",
author = "Kevin Lano and Lyan Alwakeel and Zishan Rahman",
year = "2024",
month = jul,
language = "English",
volume = "3727",
series = "CEUR Workshop Proceedings",
publisher = "CEUR-WS",
pages = "23--33",
booktitle = "STAF 2024"}

@book{Wohlin2000_ExperimentationSE,
  author = {Wohlin, Claes and Runeson, Per and H\"{o}st, Martin and Ohlsson, Magnus C. and Regnell, Bj\"{o}orn and Wessl{\'e}n, Anders},
  title = {Experimentation in software engineering: an introduction},
  year = {2000},
  isbn = {0-7923-8682-5},
  publisher = {Kluwer Academic Publishers},
  address = {Norwell, MA, USA},
}

@INPROCEEDINGS{Garcia,
  author={Garcia, Joshua and Popescu, Daniel and Edwards, George and Medvidovic, Nenad},
  booktitle={2009 13th European Conference on Software Maintenance and Reengineering}, 
  title={Identifying Architectural Bad Smells}, 
  year={2009},
  volume={},
  number={},
  pages={255-258},
  doi={10.1109/CSMR.2009.59}}

@misc{kosbar2025smellssussustainabilitysmellsiac,
      title={Smells-sus: Sustainability Smells in IaC}, 
      author={Seif Kosbar and Mohammad Hamdaqa},
      year={2025},
      eprint={2501.07676},
      archivePrefix={arXiv},
      primaryClass={cs.SE},
      url={https://arxiv.org/abs/2501.07676}, 
}

@article{Georgiou_SDL_energy_efficiency,
author = {Georgiou, Stefanos and Rizou, Stamatia and Spinellis, Diomidis},
title = {Software Development Lifecycle for Energy Efficiency: Techniques and Tools},
year = {2019},
issue_date = {July 2020},
publisher = {Association for Computing Machinery},
address = {New York, NY, USA},
volume = {52},
number = {4},
issn = {0360-0300},
url = {https://doi.org/10.1145/3337773},
doi = {10.1145/3337773},
journal = {ACM Comput. Surv.},
month = aug,
articleno = {81},
numpages = {33}
}

@INPROCEEDINGS{Durdik_guidelines,
  author={Durdik, Zoya and Klatt, Benjamin and Koziolek, Heiko and Krogmann, Klaus and Stammel, Johannes and Weiss, Roland},
  booktitle={2012 28th IEEE International Conference on Software Maintenance (ICSM)}, 
  title={Sustainability guidelines for long-living software systems}, 
  year={2012},
  volume={},
  number={},
  pages={517-526},
  doi={10.1109/ICSM.2012.6405316}}

@inproceedings{Xiao_2025,
   title={On the Effectiveness of Microservices Tactics and Patterns to Reduce Energy Consumption: An Experimental Study on Trade-Offs},
   url={http://dx.doi.org/10.1109/ICSA65012.2025.00025},
   DOI={10.1109/icsa65012.2025.00025},
   booktitle={2025 IEEE 22nd International Conference on Software Architecture (ICSA)},
   publisher={IEEE},
   author={Xiao, Xingwen and Gao, Chushu and Bogner, Justus},
   year={2025},
   month=Mar, pages={164–175} }

@INPROCEEDINGS{Albonico_robots,
  author={Albonico, Michel and Malavolta, Ivano and Pinto, Gustavo and Guzman, Emitza and Chinnappan, Katerina and Lago, Patricia},
  booktitle={2021 IEEE/ACM 18th International Conference on Mining Software Repositories (MSR)}, 
  title={Mining Energy-Related Practices in Robotics Software}, 
  year={2021},
  volume={},
  number={},
  pages={483-494},
  doi={10.1109/MSR52588.2021.00060}}

@inproceedings{Gaglianese2023_GreenCloudEdgeOrchestration,
  author       = {Marco Gaglianese and Jacopo Soldani and Stefano Forti and Antonio Brogi},
  title        = {Green Orchestration of Cloud-Edge Applications: State of the Art and Open Challenges},
  booktitle    = {{IEEE} International Conference on Service-Oriented System Engineering, {SOSE} 2023, Athens, Greece, July 17-20, 2023},
  pages        = {250--261},
  publisher    = {{IEEE}},
  year         = {2023},
  doi          = {10.1109/SOSE58276.2023.00036},
}

@article{DBLP:journals/software/VerdecchiaLEVE21,
  author       = {Verdecchia, Roberto and Lago, Patricia and Ebert, Christof and de Vries, Carol},
  title        = {Green {IT} and Green Software},
  journal      = {{IEEE} Softw.},
  volume       = {38},
  number       = {6},
  pages        = {7--15},
  year         = {2021},
  doi          = {10.1109/MS.2021.3102254}
}

@misc{GSF2021_10Recs,
    title = "10 Recommendations for Green Software Development",
    author = "San Murugesan",
    howpublished = "Green Software Foundation, \url{https://greensoftware.foundation/articles/10-recommendations-for-green-software-development}",
    year = 2021
}

@misc{GSF2021_computation,
    title = "Measure What Matters: Software's Carbon Impact",
    author = "Henry Richardson",
    howpublished = "Green Software Foundation, \url{https://greensoftware.foundation/standards/sci/}",
    year = 2026
}

@inproceedings{54_Kern2013_GreenSA,
  author={Eva Kern and Markus Dick and Stefan Naumann and Achim Guldner and Timo Johann},
  title={Green Software and Green Software Engineering – Definitions, Measurements, and Quality Aspects},
  year={2013},
  url={https://api.semanticscholar.org/CorpusID:40610952}}

@article{thematic_coding,
    author = {T. Basit},
    title = {Manual or electronic? The role of coding in qualitative data analysis},
    journal = {Educational Research, 45(2), 143–154},
    year = {2003},
    url = {https://doi.org/10.1080/0013188032000133548}
}

\end{document}